\documentclass[11pt]{article}
\pdfoutput=1

\usepackage{graphics, color,soul}
\usepackage{graphicx}
\usepackage{amssymb}

\usepackage{lmodern,mathtools}

\usepackage{booktabs}
\usepackage[english]{babel}
\usepackage{amsmath,amssymb,amsbsy,amstext, amsthm, simplewick}
\usepackage{hyperref}
\usepackage{tikz}
\usepackage{cite}

\usetikzlibrary{decorations.pathmorphing,shapes.misc}
\tikzset{snake it/.style={decorate, decoration=snake}}
\tikzset{cross/.style={cross out, draw=black, minimum size=2*(#1-\pgflinewidth), inner sep=0pt, outer sep=0pt},
cross/.default={1pt}}

\usepackage{amsfonts}
\usepackage{amssymb}
\usepackage{upgreek}
\usepackage{simplewick}
 \usepackage{exscale,relsize}
\usepackage{mathtools}
\usepackage{comment}

\usepackage[margin=1cm,labelfont={sf,bf,scriptsize},textfont={sf,scriptsize}]{caption}

\usepackage{colortbl}
\definecolor{lightgreen}{cmyk}{0.2, 0, 0.2, 0.2}
\definecolor{lightgray}{cmyk}{0.1,0.2,0,0.1}
\definecolor{lightgray2}{cmyk}{0.1,0.1,0,0.1}

\makeatletter
\newlength{\apb@width}
\newcommand{\autoparbox}[2][c]{\settowidth{\apb@width}{#2}\parbox[#1]{\apb@width}{#2}}

\makeatother

\numberwithin{equation}{section}

\def\beq{\begin{equation}}
\def\eeq{\end{equation}}

\def\bea{\begin{eqnarray}}
\def\eea{\end{eqnarray}}

\def\d{{\rm d}}

\def\beq{\begin{equation}}
\def\eeq{\end{equation}}
\def\be{\begin{equation}}
\def\ee{\end{equation}}
\def\bea{\begin{eqnarray}}
\def\eea{\end{eqnarray}}

\def\d{{\rm d}}

\def\d{{\rm d}}
\def\l{{\ell}}

\def\0{{\vec{0}}}

\def\x{{\vec{x}}}

\def\r{{\boldsymbol{r}}}

\def\x{{\bf x}}

\DeclareRobustCommand{\SkipTocEntry}[4]{}

\def\d{{\rm d}}

\def\beq{\begin{equation}}
\def\eeq{\end{equation}}
\def\d{\partial}

\def\d{\partial}
\def\l{\left(}
\def\r{\right)}

\def\ba#1\ea{\begin{align}#1\end{align}}
\def\bg#1\eg{\begin{gather}#1\end{gather}}
\newcommand{\bseq}{\begin{subequations}}
\newcommand{\eseq}{\end{subequations}}

\def\bct{{b_{CT}}}

\DeclareSymbolFont{extraup}{U}{zavm}{m}{n}
\DeclareMathSymbol{\varheart}{\mathalpha}{extraup}{86}
\DeclareMathSymbol{\vardiamond}{\mathalpha}{extraup}{87}

\def\({\left(}
\def\){\right)}
\def\[{\left[}
\def\]{\right]}

\begin{document}

\begin{titlepage}

\setcounter{page}{1} \baselineskip=15.5pt \thispagestyle{empty}

\vbox{\baselineskip14pt
}
{~~~~~~~~~~~~~~~~~~~~~~~~~~~~~~~~~~~~
~~~~~~~~~~~~~~~~~~~~~~~~~~~~~~~~~~
~~~~~~~~~~~ }

\bigskip\

\vspace{2cm}
\begin{center}
{\fontsize{19}{36}\selectfont  
{ dS/dS and $T \bar T$}
}
\end{center}

\vspace{0.6cm}

\begin{center}
Victor Gorbenko$^{1,2}$, Eva Silverstein$^2$, Gonzalo Torroba$^3$
\end{center}


\begin{center}
\vskip 8pt

\textsl{
\emph{$^1$School of Natural Sciences, Institute for Advanced Study, Princeton, 08540 NJ, USA }}
\vskip 7pt
\textsl{\emph{$^2$Stanford Institute for Theoretical Physics, Stanford University, Stanford, CA 94306, USA}}
\vskip 7pt
\textsl{ \emph{$^3$Centro At\'omico Bariloche and CONICET, Bariloche, Argentina}}

\end{center}

\vspace{0.5cm}
\hrule \vspace{0.1cm}
\vspace{0.2cm}
{ \noindent \textbf{Abstract}
\vspace{0.3cm}

The $T\bar T$ deformation of a conformal field theory has a dual description as a cutoff $AdS_3$ spacetime, at least at the level of pure 3d gravity.  We generalize this deformation in such a way that it builds up a patch of bulk $dS_3$ spacetime instead.  At each step along the trajectory in the space of $2d$ theories, the theory is deformed by a specific combination of $T\bar T$ and the two-dimensional cosmological constant.   This provides a concrete holographic dual for the warped throat on the gravity side of the dS/dS duality, at leading order in large central charge.   We also analyze a sequence of excitations of this throat on both sides of the duality, as well as the entanglement entropy. Our results point toward a mechanism for obtaining de Sitter solutions starting from seed conformal field theories with AdS duals.

\vspace{0.4cm}

 \hrule

\vspace{0.6cm}}
\end{titlepage}

\tableofcontents

\section{Introduction}\label{sec:intro}

The problem of formulating quantum gravity in generic, cosmological spacetime remains of fundamental interest.  Among the approaches to holography for de Sitter and its decays \cite{Anninos:2012qw}\cite{Strominger:2001pn, Strominger:2001gp}\cite{Alishahiha:2004md, Alishahiha:2005dj}\cite{Freivogel:2006xu}\cite{Dong:2010pm, Dong:2011uf} \cite{Anninos:2011ui} , the dS/dS correspondence \cite{Alishahiha:2004md, Alishahiha:2005dj} proceeds by uplifting the AdS/CFT correspondence as described in \cite{Dong:2010pm, Dong:2011uf}.  As such it retains some common features such as an emergent spatial direction and unitarity of the dual Lorentzian theory, a highly constrained holographic RG flow \cite{Dong:2012afa}, and well-defined Von Neumann and Renyi entropies which provide a clear interpretation of the de Sitter entropy as entanglement entropy \cite{Dong:2018cuv}.\footnote{Other interesting works such as \cite{Miyaji:2015yva}\ and \cite{Nomura:2018kji}\ also study entanglement entropy in finite patches of spacetime.}       

These results build on the macroscopic observation that de Sitter is a warped compactification to lower-dimensional de Sitter, e.g.
\be\label{dSdSmetric}
ds^2_{(A)dS_{3}}=dw^2+{\text{sin(h)}}^2\frac{w}{\ell} ds^2_{dS_2}=dw^2+{\text{sin(h)}}^2\frac{w}{\ell}(-d\tau^2+\ell^2\cosh^2(\frac{\tau}{\ell})d\phi^2)
\ee
The bulk dS and bulk AdS theories agree in the highly redshifted region ${\text{sin(h)}}^2\frac{w}{\ell}\ll 1$,
indicating that the dual of dS will involve some kind of irrelevant deformation relative to the CFT dual of AdS.
The most ultraviolet scale in the dS/dS theory is finite, $\sqrt{g_{00}}=\sin\frac{w_{uv}}{\ell}=1$, indicating a cutoff matter theory constrained by $2d$ gravity.  The statement that this matter theory consists of two isomorphic sectors constrained by gravity follows independently from the uplift of the AdS/CFT brane construction \cite{Dong:2010pm}, and a similar formulation is also ultimately required for the dS/CFT conjecture as well \cite{Maldacena:2002vr, Harlow:2011ke}.  These features strongly motivate us to investigate the duality further.  
Extensive calculations in the works referenced above (and others) provide concrete clues as to the nature of this dual, but its precise formulation is not known.  

Enter the $T\bar T$ deformation:  recently Zamolodchikov and Smirnov \cite{Zamolodchikov:2004ce, Smirnov:2016lqw}\ and Cavaglia et al \cite{Cavaglia:2016oda} introduced a powerful prescription for controlling a particular irrelevant deformation in two dimensions.    Starting from any seed quantum field theory, they specify a trajectory in the space of $2d$ theories along which exact calculations of energy levels and other quantities remain tractable, at least on flat spacetime.   There has been extensive analysis and discussion in the literature of the interpretations and applications of this~\cite{Dubovsky:2012wk, McGough:2016lol,  Dubovsky:2017cnj, Cardy:2018sdv, Cottrell:2018skz, Kraus:2018xrn, Donnelly:2018bef, Hartman:2018tkw, Taylor:2018xcy, Bonelli:2018kik, Aharony:2018vux, Aharony:2018bad, Datta:2018thy, Giveon:2017nie, Guica:2017lia, Baggio:2018rpv, Chang:2018dge}.

We will build from the observation of McGough, Mezei, and Verlinde \cite{McGough:2016lol}\ that the dressed energy levels and other calculable quantities under the $T\bar T$ deformation agree precisely with those of a cut off $AdS_3$ spacetime.   Since it involves an irrelevant deformation, moving onto the $T\bar T$ trajectory has a large effect on the would-be ultraviolet region of the geometry in the holographic dual.  The work \cite{McGough:2016lol}\ argued that the effect is simply to chop it off.  At the level of pure gravity, this was confirmed and extended in several ways in the works \cite{Kraus:2018xrn, Donnelly:2018bef}, which also noted that beyond pure gravity the field theory side prescription may become more involved in order to capture effects of bulk matter and cut them off.  As we will see in this paper, the field theory side of the duality can be adjusted to account for new features of the bulk, such as $dS$ rather than $AdS$ geometry.  We will leave the QFT-side treatment of additional bulk matter fields to future work, but note that our method for generalizing the trajectory may extend to those cases \cite{usnext}. This may also relate to the work \cite{Giveon:2017nie} involving a radially varying scalar field (string dilaton).   A similar question arises for string theory even at the classical level; we will discuss an approach to this in the conclusions.     

The duality conjecture \cite{McGough:2016lol}\ relates the quasilocal stress-energy tensor to the deformed QFT stress-energy tensor, generalizing \cite{Balasubramanian:1999re}.  The bulk Einstein equations imply that the former satisfies the differential equations derived for the latter in \cite{Cavaglia:2016oda, Smirnov:2016lqw}.  Moreover,  the large-$c$ factorization of correlation functions simplifies the field-theory side analysis, a feature that is useful for generalizations to curved 2d spacetimes as in \cite{Donnelly:2018bef}\ and the higher-dimensional generalization in \cite{Hartman:2018tkw, Taylor:2018xcy}.\footnote{The recent literature includes numerous contributions to the interpretation and generalization of the $T\bar T$ deformation. { For example, a derivation of the equivalence of the $T\bar T$ deformation to a certain version of Jackiw-Teitelboim gravity appeared in  \cite{Dubovsky:2017cnj} for the Minkowski space and in  \cite{Dubovsky:2018bmo} for the torus; other interesting works which included a relation between $T\bar T$ and the metric fluctuations appeared in \cite{Cardy:2018sdv, Cottrell:2018skz}.}  Closed-form expressions for the action along the trajectory for a variety of seed theories appeared in \cite{Bonelli:2018kik}. An interesting study of the large-$c$ limit appeared in \cite{Aharony:2018vux}, and of modular transformations in \cite{Aharony:2018bad, Datta:2018thy}. Some generalizations of the $T\bar T$ deformation appeared in \cite{Giveon:2017nie, Guica:2017lia}.}  

Given all this, it is natural to investigate, even at the level of pure large-c gravity, whether a generalization of the $T\bar T$ deformation could generate the dS/dS cut off warped throat.  In this work, we find that this is indeed the case.   Applying the same gravity-side analysis of \cite{McGough:2016lol, Kraus:2018xrn}\ to the $dS_3$ warped throat, we obtain its quasilocal stress-energy tensor and an algebraic equation it satisfies, generalizing the AdS/CFT case to our case of positive bulk cosmological constant.   The resulting defining equation is quite simple, containing one extra term compared to the cutoff AdS/CFT case.  We show how this term arises on the other side of the duality if we incorporate a relevant deformation along with the $T\bar T$ deformation at each step in the trajectory in the space of $2d$ theories.   This arises from a simple generalization of the differential equation derived in  \cite{Cavaglia:2016oda, Smirnov:2016lqw}.  We obtain a boundary condition on the trajectory for the bulk $dS_3$ case by matching the behavior of AdS/dS (i.e. the CFT on $dS_2$) in the regime where they coincide, i.e. small $\text{sin(h)}(w/\ell)$ (\ref{dSdSmetric}).  This trajectory and boundary condition yields a dual $2d$ construction of the dS/dS warped throat, and a family of particle excitations of it, starting from any seed CFT.
Moreover, we find that the perturbations in our cutoff theory do not exhibit the superluminal behavior that arose in the pure $T\bar T$ deformations \cite{Zamolodchikov:2004ce, Smirnov:2016lqw} with the sign corresponding to the dual cutoff AdS \cite{McGough:2016lol, Kraus:2018xrn}.  This is related to the absence of horizons for massive particles in $dS_3$.  However, we have not analyzed more general excitations of the system or their perturbations.  We also analyze the entanglement entropy obtained by tracing out half of the space on which the $2d$ theory lives, as a function of the $dS_2$ radius $r$, and find precise agreement between the two sides of the proposed duality.  

This paper is organized as follows.  
Section 2 contains the gravity-side analysis of the quasilocal stress-energy tensor and Einstein equations, including excitations introduced by massive particles.  Section 3 derives the generalized $T\bar T$ deformation of the 2d dual that reproduces the gravity results. Along the way, we clarify some aspects of the $T \bar T$ deformation in curved space, including the role of the Weyl anomaly. In section 4, we compute the entanglement entropy on both sides.  Section 5 contains a summary and discussion of future directions. Some more detailed calculations are presented in the Appendix.

\section{Trajectory from the (A)$dS_3$ bulk}\label{sec:bulk}

In this section, we will analyze the quasilocal stress-energy tensor on the $AdS_3$ and $dS_3$ warped
throats (\ref{dSdSmetric}) and their excitations, with a radial cutoff at a finite value $w=w_c$.  This will generalize the bulk $AdS_3$ treatments in \cite{McGough:2016lol, Kraus:2018xrn, Donnelly:2018bef}.  It will provide a simple generalization of the equations for the dressed energy levels derived on the bulk gravity side.  

Then in \S\ref{sec:dual}, we will construct an explicit generalization of the $T\bar T$ trajectory that reproduces this flow.   This will reconstruct the bulk $dS_3$ quasilocal stress-energy, which is closely related to the bulk $dS_3$ geometry, starting from the field theory side.    

In this analysis, we consider each warped throat of dS/dS separately; i.e. we work with $0\le w_c\le \pi/2$.
Each of the two warped throats in  (\ref{dSdSmetric}) is a building block for the dS/dS correspondence, one which we will construct from the field theory side in this paper.  It will be interesting to analyze the dual description of their coupling in future work.

\subsection{The trace flow equation}\label{subsec:traceflowGR}

We will start our gravity-side analysis by deriving the generalization of a basic equation relating the trace of the stress-energy tensor to $T\bar T$, along the lines of
\cite{McGough:2016lol, Kraus:2018xrn, Donnelly:2018bef}.
It is straightforward to generalize the derivation in~\cite{Kraus:2018xrn}\ to the bulk $dS_3$ case.  The Einstein-Hilbert plus Gibbons-Hawking-York action is
\be\label{action}
S= \frac{1}{16\pi G}\,\int_{\mathcal M}\,d^3 x\,\sqrt{-g} \left(\mathcal R^{(3)} + \frac{2 \eta}{\ell^2} \right)+ \frac{1}{8\pi G}\,\int_{\partial \mathcal M}\,d^2 x\,\sqrt{-g} \left(K -\frac{b_{CT}}{\ell} \right)\,,
\ee 
with $\eta=+1$ for AdS and $\eta=-1$ for dS.  Here we note the counterterm $\bct$, which was set to 1 in \cite{Kraus:2018xrn}.  We will not carry $\bct$ through our analysis, setting it to 1 in order to focus on the effect of flipping the sign of $\eta$, holding all else fixed.\footnote{For completness, in App. \ref{subsec:appflow} we determine the effect of $b_{CT}$ on the flow equation.}  It is also possible to add a boundary Wess-Zumino term when the boundary metric is curved as in (\ref{dSdSmetric}). This, however, will not play an important role in our present discussion, so we will set it to zero. The possibility of a more general boundary action will be interesting to investigate more systematically in the future, in the process of joining our warped throats together.  

Taking radial slices
\be\label{eq:radial}
ds_3^2 = dw^2 + g_{ij}\,dx^i dx^j\,,
\ee
the quasilocal stress-tensor becomes\footnote{Our convention for the stess-tensor is the same as in~\cite{Balasubramanian:1999re}, which differs from~\cite{Kraus:2018xrn} by a factor of $1/(2\pi)$.}
\be\label{bCTT}
T_{ij}=\frac{2}{\sqrt{-g}}\frac{\delta S_\text{on-shell}}{\delta g^{ij}}=\frac{1}{8\pi G} \left(K_{ij}-K g_{ij}+\frac{b_{CT}}{\ell}g_{ij}\right)\,.
\ee
In the coordinates (\ref{eq:radial}), the extrinsic curvature is $K_{ij}= \frac{1}{2}\partial_w g_{ij}$.
We will keep track of $\eta$ throughout our analysis, our main goal being to understand the case $\eta=-1$ corresponding to the $dS_3$ bulk.

Applying the Einstein equation $E^w_w=0$ following \cite{Kraus:2018xrn}, with $E^\mu_\nu$ the Einstein tensor, we find a more general trace flow equation:
\be\label{TequationdS}
T^i_i =-\frac{\ell}{ 16\pi G}\,{\cal R}^{(2)}-4\pi G\ell\,\left(T^{ij} T_{ij}-(T^i_i)^2 \right)-\frac{\eta-1}{8\pi G \ell}\,,
\ee 
where $\mathcal R^{(2)}$ is the Ricci scalar of $g_{ij}$ at fixed $w$. The calculation is presented in Appendix \ref{app:GReqs}. This
generalizes equation (3.7) in~\cite{Kraus:2018xrn}. The $AdS_3$ result with a flat cylindrical boundary is recovered if we turned off ${\cal R}^{(2)}$, take $\eta=1$ (and as already noted, rescale $T_{ij}$ to work in the conventions \cite{Balasubramanian:1999re}).  Of course gravity solutions must obey the full set of Einstein equations with appropriate boundary conditions, which we will incorporate below and in the appendix.  

We will perform a detailed comparison with the two-dimensional dual that we construct in \S \ref{sec:dual}, with the map of parameters collected in \S\ref{subsec:parameters}. For now, let us anticipate some elements of this dictionary starting from the field theory result perturbed by $T\bar T$:
\be\label{QFTTrT}
T^i_i =-\frac{c}{24\pi} \mathcal R^{(2)}  -4 \pi \lambda T \bar T+ \ldots 
\ee
where we used the notation 
\be\label{eq:TTbdef}
T\bar T=\frac{1}{8}(T^{ij} T_{ij}-(T^i_i)^2)\,.
\ee  
In the single-scale case treated in \cite{McGough:2016lol, Kraus:2018xrn}, this corresponds to a deformation
generated at the level of the action by
\be\label{TTbarAction}
\frac{dS}{d\lambda}=2\pi \,\int d^2 x\,\sqrt{-g}\,T\bar T
\ee     
since in that case $\int d^2 x\,\sqrt{-g}\,T^i_i=\mu dS/d\mu$ and $\mu = 1/\lambda^{1/2}$.\footnote{The scaling dimension of the coupling $\lambda$ equals $-2$ along the field theory trajectory, because in the large $N$ limit $T \bar T$ factorizes, and the stress tensor always has scaling dimension $2$. At finite $N$, this also holds for a theory in flat space \cite{Zamolodchikov:2004ce}.}        
This allows us to identify~\cite{McGough:2016lol}
\be\label{eq:identify}
c=\frac{3 \ell}{2G}\;,\;\lambda = 8 G \ell\;\Rightarrow\; \lambda c = 12 \,\ell^2\,.
\ee
Therefore, in the large central charge limit, with radius $\ell$ fixed, the deformation parameter scales as $\lambda \sim 1/c$. 

Altogether, we obtain 
\be\label{flowQFTvar}
T^i_i = -\frac{c}{24\pi} {\cal R}^{(2)} - 4\pi \lambda T\bar T -\frac{{\eta}-1}{\pi\lambda}\,,
\ee
which we will refer to as the trace flow equation.  
This contains an additional term compared to the case of bulk AdS \cite{Donnelly:2018bef}.  Our goal in \S\ref{sec:dual}\ will be to obtain this term, and a related one in the flow of energy levels, from a generalization of the field-theoretic flow defined in \cite{Cavaglia:2016oda, Smirnov:2016lqw}.

\subsection{Solution for the dS ground state}\label{subsec:ground-state}

For the de Sitter invariant ground state, the stress tensor should be of the form
$T_{ij}=\alpha g_{ij}$; similarly to \cite{Donnelly:2018bef}, from (\ref{flowQFTvar}) we obtain the quadratic equation
\be\label{alphaeq}
8 \pi G\alpha^2-\frac{2}{\ell}\alpha - \frac{\eta-1}{8\pi G \ell^2}-\frac{{\cal R}^{(2)}}{16\pi G} = 0,
\ee
which gives
\be\label{alphaus}
\alpha = \frac{1}{4\pi G \ell}\left(\frac{1}{2}\pm \sqrt{\frac{{\cal R}^{(2)}}{8}+\frac{\eta}{4\ell^2}} \right)\,.
\ee
where again $\eta=+ 1$ ($\eta=-1$) is for the AdS (dS) bulk case. 
For a slicing of (A)$dS_3$ by $dS_2$ with
\be
\mathcal R^{(2)}= \frac{2}{r^2}
\ee
with $r$ the curvature radius of the boundary,
we obtain
\be\label{alpha2}
\alpha = \frac{1}{8\pi G \ell}\left(1 \pm \sqrt{\frac{\ell^2}{r^2}+\eta} \,\right)\,,
\ee
exhibiting the significant effect of $\eta$.  

Before coming to the interpretation of this, we need to impose a boundary condition on the flow in order to choose between the $\pm$ branches.
For AdS with $\eta=1$ we can do this by analyzing the UV limit arising as we approach the boundary, $\lambda/r^2 \to 0$. The expectation value of the stress-tensor should be given solely by the Weyl anomaly (the last term in (\ref{TequationdS})). This defines a UV boundary condition given by the minus branch:
\be\label{eq:alphaAdS}
\alpha_\text{AdS} = \frac{1}{8\pi G \ell}\left(1 - \sqrt{\frac{\ell^2}{r^2}+1} \,\right)=\frac{1}{\pi \lambda}\left(1-\sqrt{\frac{\lambda c}{12 r^2}+1}\, \right)\,.
\ee
as derived earlier in \cite{Donnelly:2018bef}.

On the other hand, for a dS bulk we have $\eta=-1$,
\be
\alpha = \frac{1}{8\pi G \ell}\left(1 \pm \sqrt{\frac{\ell^2}{r^2}-1} \,\right)\,.
\ee
This imposes an upper bound $r < \ell$, correctly for dS/dS, where the warp factor is $a(w)= \ell \sin (w/\ell)$. We can no longer impose the boundary condition for the RG flow in the UV limit $\lambda/r^2 \to 0$. Instead, we note that the opposite limit of a large irrelevant deformation, $\lambda/r^2 \to \infty$, should impose a radial cutoff that is approaching the IR part of the geometry. In this limit, the bulk AdS and bulk dS throats agree \cite{Alishahiha:2004md}. Therefore, we will fix the boundary condition for the dS/dS flow by matching to the AdS behavior at $\lambda/r^2 \to \infty$, and this sets
\be\label{eq:alphadS}
\alpha_\text{dS} = \frac{1}{8\pi G \ell}\left(1 - \sqrt{\frac{\ell^2}{r^2}-1} \,\right)\, = \frac{1}{\pi \lambda}\left(1-\sqrt{\frac{\lambda c}{12 r^2}-1}\, \right)\,.
\ee
Here we also use this limit to identify the central charge,  obtaining the same expression as in AdS, $c=\frac{3\ell}{2G}$. 

We note that the choice of the `$-$' branch agrees with the result from (\ref{bCTT}) for the quasi-local stress-tensor. See \S \ref{subsec:reconstruct} below for more details.

\subsection{The dressing of higher energy levels}\label{subsec:excited}

We would also like to understand the higher energy levels in the theory.   For negative bulk cosmological constant ($\eta=1$), there are particle states (corresponding to a conical deficit angle) and black hole solutions which asymptote to $AdS_3$.  For positive bulk cosmological constant,  the massive particles produce a simple identification of points on the original $dS_3$ \cite{Deser:1983nh, Bousso:2001mw}.  For example for vanishing spatial momentum, $p_\phi=0$, the effect is to change the periodicity in the $\phi$ direction: 
\be\label{dSBH}
ds^2_{3\mu}=dw^2 + \sin^2(\frac{w}{\ell}) (-d\tau^2+\ell^2\cosh^2\frac{\tau}{\ell} d\phi^2)\;\;,\;\;\phi=\phi+2\pi \mu
\ee
with $\mu$ related to the mass $m$ of the particle as 
\be\label{mumass}
m=\frac{1}{4 G}\frac{\Delta\phi}{2\pi}=\frac{1}{4G}(1-\mu)\,.
\ee
The proper size of the circle that the $2d$ boundary theory lives on is 
\be\label{Lr}
L=\int d\phi\sqrt{g_{\phi\phi}} =2\pi \mu\,\ell\,\sin(\frac{w_c}{\ell})\,\cosh(\frac{\tau}{\ell})\,.
\ee    
Note that this size shrinks with reduced $\phi$ period, but the curvature radius remains the same.  

Our goal in this paper is to reconstruct the $3d$ de Sitter ground state and this family of excited particle solutions, from a field theory flow involving $T\bar T$.  There are various ways we could set that up.  The simplest way to capture the energy levels will be to work with the QFT on a `tall de Sitter' spacetime obtained from the original $dS_2$ by reducing the periodicity of $\phi$ to $2\pi\mu$.  
\be\label{dS2tall}
ds^2_{2\mu}=\left(\text{sin(h)}(w_c/\ell)\right)^2\left(-d\tau^2+\ell^2\cosh^2\frac{\tau}{\ell} d\phi^2\right), ~~~~~\phi=\phi+2\pi n\mu\,.
\ee
It will prove convenient to work with this as the spacetime on which the QFT lives, reconstructing the bulk solution in a given mass sector (\ref{mumass}) from the dressed {\it ground state} of the QFT on the corresponding tall de Sitter spacetime.  
This allows us to directly map the QFT calculation to the gravity side with the boundary at fixed $w=w_c$ in the metric (\ref{dSBH}).\footnote{It is also possible to keep the boundary geometry completely fixed, as we discuss briefly below in \S\ref{generalconjecture}\ in the context of more general excitations of the system.}

These excited levels are in fact inherited from the above solution for the vacuum.  For all $\mu$, we can use two algebraic equations to obtain the dressed stress-energy. If we first apply the trace flow equation (\ref{flowQFTvar}), we can solve for $T^\phi_\phi$ algebraically in terms of $T^\tau_\tau$.  This gives
\be\label{TphiTtautext}
T^\phi_\phi = \frac{T^\tau_\tau+c \frac{{{\cal R}^{(2)}}}{24\pi} +\frac{\eta-1}{\pi\lambda}}{\pi\lambda T^\tau_\tau-1}\,.
\ee
If we are in the ground state in $dS_2$, or an orbifold of that (as in the tall $dS_{2\mu}$ spacetimes), we have another algebraic relation:  $T^\phi_\phi = T^\tau_\tau\equiv \alpha$.  This relation and (\ref{TphiTtautext}) together yield the solution for $\alpha$ given above, namely
\be\label{Ttautauanswertext}
T^\tau_\tau =\frac{1}{\pi\lambda}\left(1-\sqrt{\eta+ \frac{c\lambda {\cal R}^{(2)}}{24}} \right)= \frac{1}{\pi \lambda}\left(1-\sqrt{\frac{\lambda c (2\pi\mu)^2\cosh^2(\tau/\ell) }{12 L(\tau)^2}+{\eta}} \right)\,.
\ee
In the second expression, we wrote this in terms of $L$ and the period $\mu$ of the circle, which is related to the particle source mass $m$ above as in (\ref{mumass}).   

It is interesting to translate $\mu$ to field theory language, as reviewed e.g. in \cite{Martinec:1998wm}.  The particle states with mass sourcing a deficit angle (\ref{mumass}) correspond to dimension
\be\label{Lzero}
L_0=\tilde L_0 = \frac{c}{12}\mu(1-\mu)
\ee 
and ADM mass $\ell M= L_0 + \tilde L_0 +c/12$.  

In the appendix, we collect other gravity-side relations relevant for characterizing the bulk geometry and stress-energy tensor.  These may also be useful for a future study of more general excitations.  
In \S\ref{sec:dual}, we will derive the new term proportional to $\eta-1$ in the flow equation (\ref{TequationdS}), which arose from the positive bulk cosmological constant, by defining a generalized flow involving both $T\bar T$ and a relevant deformation.

\subsection{Luminality of the perturbations}\label{subsec:luminality}

Let us next explore the speed of propagation of `boundary graviton' perturbations in our bulk-$dS_3$ warped throat.  This is an interesting issue at the level of a single warped throat with the Dirichlet cutoff, the building block for the dS/dS correspondence that we are formulating in this work.   However, we note that once the throats join to produce the full (two-throat) $dS_3$ system, we will not have these boundary modes.     
   
We will follow the steps derived in \cite{Marolf:2012dr}\ for the $AdS_3$ case, applying it to our system with Dirichlet boundary conditions at the cutoff surface $w=w_c$ in (\ref{dSBH}).   There is a substantial difference between the bulk AdS and bulk dS cases:  as noted above, the particles in $dS_3$ do not introduce a black hole horizon.  In \cite{Marolf:2012dr}\ it was the emblackening factor of the BTZ black hole that led to superluminal perturbations of boundary gravitons, matching those that had been identified in the $T\bar T$ deformation on a cylinder, as reviewed in \cite{McGough:2016lol, Kraus:2018xrn}.  As we will see shortly, the analogous modes in our system are not superluminal, simply because of this difference between particle states and black holes.\footnote{One can also think about the problem in terms of signals sent through the bulk versus those staying on the boundary.  The $dS_2$ boundary plays a positive role in that signals sent into the bulk take longer compared to the cylinder case to reach the boundary, if they do at all.  We thank XiaoLiang Qi and Juan Maldacena for discussions of this point.}       

We start from the form (\ref{metric}) of the metric, perturbing around the fiducial solution (\ref{dSBH}) with $g=\sin (w/\ell), r=\ell\cosh(\tau/\ell)\sin (w/\ell)$.  Specifically, we introduce small perturbations of the cutoff surface:  $w_c\to w_c+\delta w(\phi,\tau)$, requiring them to preserve the curvature of the metric as required by the Dirichlet problem at our boundary.  
This is equivalent to the original Dirichlet problem.\footnote{We thank Mukund Rangamani for discussions of this point.} We could have started with a metric perturbation in the bulk which vanishes at the boundary.  Performing a coordinate transformation which removes the perturbation in the bulk will generally not vanish at the boundary.  This surviving mode is $\delta w(\tau, \phi)$.  Such a procedure will automatically preserve the curvature invariant ${ \cal R}^{(2)}$ at the boundary.  Thus, in order to capture the correct deformations we must impose the condition $\delta { \cal R}^{(2)}=0$.

Let us start from the metric in the form given in the Appendix, equation (\ref{metric}).  
With this perturbation, the induced metric on the boundary becomes (at the linearized level)
\be\label{pertmetric}
ds^2_2\to -g^2(1+2\frac{\partial_w g}{g}\delta w)d\tau^2+r^2(1+2\frac{\partial_w r}{r} \delta w)d\phi^2\simeq e^{2\delta w (\partial_w r/r)|_{w_c}}ds^2_2
\ee
where we used the fact that $\partial_w g/g = \partial_w r/r = \frac{\cos(w/\ell)}{\ell\sin(w/\ell)}$.  

Under a Weyl transformation $g_{\mu\nu}\to e^{2\omega}g_{\mu\nu}$, the scalar curvature transforms as ${\cal R}^{(2)}\to {\cal R}^{(2)}(1-2\omega)-2\nabla^2\omega $.  Setting this equal to the original scalar curvature implies
\be\label{curvatureequal}
\nabla^2\omega + {\cal R}^{(2)}\omega =0\;,\;\;\; \omega=\delta w \frac{\cos(w/\ell)}{\ell\sin(w/\ell)}\,.
\ee
For a generic value of $w_c$, this is an equation for a negative mass-squared scalar on the (tall) de Sitter spacetime.  It is not acausal, but apparently does indicate a Hubble scale instability.  Interestingly, when we take $w_c=\pi\ell/2$, the most ultraviolet slice of the dS/dS correspondence, the coefficient $\frac{\cos(w/\ell)}{\ell\sin(w/\ell)}$ vanishes, relegating the question to higher orders.  In any case, we see here that the bulk dS case lacks the superluminal boundary gravitons that appeared in the original examples (for one sign of $\lambda$ in those cases).   Again, this makes sense given the absence of a black hole horizon in our case.   

One could go further and analyze the action for this perturbation.  However, our analysis in this paper is classical (large-$c$ in the $d$-dimensional dual), and as mentioned above the joining of the two throats to obtain the full $dS_3$ system will not contain these modes.  For these reasons, we will defer additional discussion of boundary conditions to future work on the joined system.    

\subsection{Reconstructing the bulk geometry}\label{subsec:reconstruct}

As a final remark in our analysis of the gravity side, we point out that the flow equation for the quasi-local stress tensor gives a way to reconstruct the bulk geometry. Let us write the bulk metric in terms of a warp factor $A(w)$,
\be
ds^2= dw^2 +  g_{ij}(w,x) \,dx^i dx^j=dw^2 + e^{2A(w)} \hat g_{ij}(x)\, dx^i dx^j\,,
\ee
with $\hat g_{ij}$ a constant curvature metric independent of $w$.
At a radial position $w=w_c$, the stress tensor (\ref{bCTT}) becomes
\be
T_{ij}= \frac{1}{8 \pi G \ell}\left(1 - \ell \partial_w A \right) g_{ij}\,.
\ee
On the other hand, the solution (\ref{alphaus}) to the flow equation gives
\be\label{eq:TA}
T_{ij}= \frac{1}{8 \pi G \ell}\left(1 - \sqrt{ e^{-2A(w)}+ \eta} \right) g_{ij}\,,
\ee
where for simplicity we take the curvature of $g_{ij}$ to be
\be
\mathcal R^{(2)}= \frac{2}{\ell^2} \,e^{-2A(w)}\,.
\ee
This corresponds to a metric $\hat g_{ij}$ of scalar curvature $2/\ell^2$.

Combining both equations gives a differential equation for the warp factor,
\be
(\partial_w A)^2 - e^{-2A(w)} - \frac{\eta}{\ell^2}=0\,,
\ee
whose solution
\be
e^{2A(w)} = \left( \frac{1}{\sqrt{\eta}}\,\sinh\left( \sqrt{\eta}\frac{w}{\ell}\right)\right)^2
\ee
reproduces the bulk (A)dS geometry for $\eta= \pm 1$. More general solutions are discussed in App. \ref{subsec:appexcited}. In the next section we will derive the trace flow equation directly in the $2d$ dual; a relation like (\ref{eq:TA}) allows then to reconstruct the bulk geometry directly in terms of field-theoretic quantities.

\section{$T \bar T$ and $\Lambda_2$ flow in the $2d$ dual}\label{sec:dual}

Now that we know what our bulk theory does in the class of states discussed in \S\ref{sec:bulk}\ (for each throat), let us analyze the $T \bar T$ trajectory directly on the two-dimensional dual. The factorization of $T \bar T$ was proved at finite central charge in~\cite{Zamolodchikov:2004ce}, when the space is a plane or a cylinder. It is not clear yet if this holds in more general curved spaces, and understanding this would be an interesting direction. Here we will focus on the large-$c$ limit, where correlators automatically factorize. 


\subsection{Defining the $2d$ trajectory}\label{subsec:qftflow}

Let us consider a $2d$ theory living on a spacetime whose metric is the induced metric on the boundary at $w=w_c$. 
In our system, this is is the spacetime (\ref{dS2tall}).   It depends on the mass sector, related to the parameter $\mu$, that we wish to reconstruct.     We would like to define a trajectory in the space of $2d$ theories, generalizing  \cite{Cavaglia:2016oda, Smirnov:2016lqw}, which captures the new term in the trace flow equation (\ref{flowQFTvar}) arising on the gravity side.     

In the recent work \cite{Donnelly:2018bef}, the trace flow equation (\ref{flowQFTvar}) with $\eta=1$ was simply treated as a defining equation of the trajectory in the space of $2d$ theories.   
We will shortly obtain this relation via a step by step specification of the trajectory in the space of couplings in the $2d$ theory.  One can then apply it on homogeneous states to solve algebraically for $T^\phi_\phi$ as described above in (\ref{TphiTtautext}), leading to dressed stress-energy (\ref{Ttautauanswertext}) for the vacuum and simple particle excitations.

Before proceeding to our generalization, let us first explain how this algebraic manipulation fits with the original dressed energy levels in  \cite{Cavaglia:2016oda, Smirnov:2016lqw}, working for simplicity at zero spatial momentum.  Applying the trace flow equation in homogeneous states, we can solve for the expectation value $\langle T^\phi_\phi \rangle$ in terms of $\langle T^\tau_\tau \rangle$ (turning off the curvature and $\eta-1$ terms in (\ref{TphiTtautext})).  Then identifying $\langle T^\phi_\phi \rangle $ with the pressure and solving the resulting differential equation we recover the standard result for the dressed energy levels:
\bea\label{TFEpressureTalk}
{\rm Tr}\, T&=&T^\tau_\tau+T^\phi_\phi = -4\pi \lambda T\bar T \Rightarrow \langle T^\phi_\phi\rangle = \frac{\langle T^\tau_\tau \rangle}{\pi\lambda \langle T^\tau_\tau \rangle-1}\nonumber\\
\langle T^\phi_\phi \rangle &=& -d E/dL\nonumber\\
\langle T^\tau_\tau \rangle &=&-\frac{E}{L}=\frac{1}{\pi\lambda}\left(1-\sqrt{1+ 2\pi\lambda \langle {T^\tau_\tau}^{(0)}\rangle } \right)
\eea            
where $\langle{T^\tau_\tau}^{(0)}\rangle=-E^{(0)}/L$ is (minus) the undressed energy density. Note that for the case of the deformed CFT one can use the trace flow equation to define the trajectory, instead of the operator relation used in the original works. Consequently, on the cylinder these approaches are equivalent. 

We will now apply the same strategy as in  \cite{Cavaglia:2016oda, Smirnov:2016lqw} to the theory on $dS_2$, which in Euclidean signature becomes a two-sphere 
\be\label{eq:S2}
ds^2=r^2 (d\theta^2+\sin^2\theta d\phi^2)\,.
\ee 
We start with the pure $T\bar T$ deformation which is expected to match the  $AdS_3$ bulk. Let us start from a symmetric state, in which 
\be
\langle T^\theta_\theta \rangle=\langle T^\phi_\phi \rangle= \frac{1}{2}\,\langle {\rm Tr}\,T \rangle
\ee
 is independent of the position on the sphere. Then the $T\bar T$ deformation gives, upon integrating over the volume of the sphere,
\be\label{eq:ZTTbnew}
\frac{\partial}{\partial \lambda}\, \log Z =-2\pi \,\int d^2 x\,\sqrt{g}\, \langle T \bar T \rangle=-8\pi^2r^2 \langle T\bar T\rangle\,.
\ee
On the other hand, 
\be\label{eq:Trace}
r \frac{\partial}{\partial r}\, \log Z =- \int d^2 x\,\sqrt{g}\,\langle \text{Tr} T \rangle=-8\pi r^2\,\langle T^{\theta}_\theta \rangle\,.
\ee
According to \eqref{eq:TTbdef}, for a symmetric state we also have 
\be
\langle T\bar T \rangle=-\frac{1}{4}\langle T^{\theta}_\theta \rangle^2\,,
\ee
where in this last step we used large-$c$ factorization.
We then differentiate \eqref{eq:ZTTbnew} with respect to $r$ to get
\be
-\frac{\d}{\d\lambda} 8\pi r\,\langle T^{\theta}_\theta \rangle= \frac{\d}{\d r}\l 2\pi^2 r^2  \,\langle T^{\theta}_\theta \rangle^2 \r\, ,
\ee
or if, to compare with the zero curvature case, we introduce $E=- 2\pi r \,\langle T^{\theta}_\theta \rangle$, 
\be
\label{eq:SZ}
4\, \frac{\d}{\d\lambda}\, E(r,\lambda) = E(r,\lambda)\, \frac{\d}{\d r} \,E (r,\lambda)\;.
\ee
Interestingly, the energy on the equator of the sphere satisfies exactly the same ``hydrodynamic'' equation, as the energy levels on the cylinder do, cf  \cite{Cavaglia:2016oda, Smirnov:2016lqw}.  In Lorentzian signature, this becomes the neck of the $dS_2$, where the Christoffel symbols vanish and stress-energy conservation takes the same form as on the cylinder.  If the value of energy is known for $\lambda=0$, \eqref{eq:SZ} can be readily solved for a finite $\lambda$. In particular, if the undeformed theory is conformal, the solution for $\langle T^{\theta}_\theta \rangle $ agrees with \eqref{eq:alphaAdS}. Note that \eqref{eq:SZ} applies equally well if the undeformed theory is massive.

\begin{figure}[htbp]
\begin{center}
\includegraphics[width=12cm]{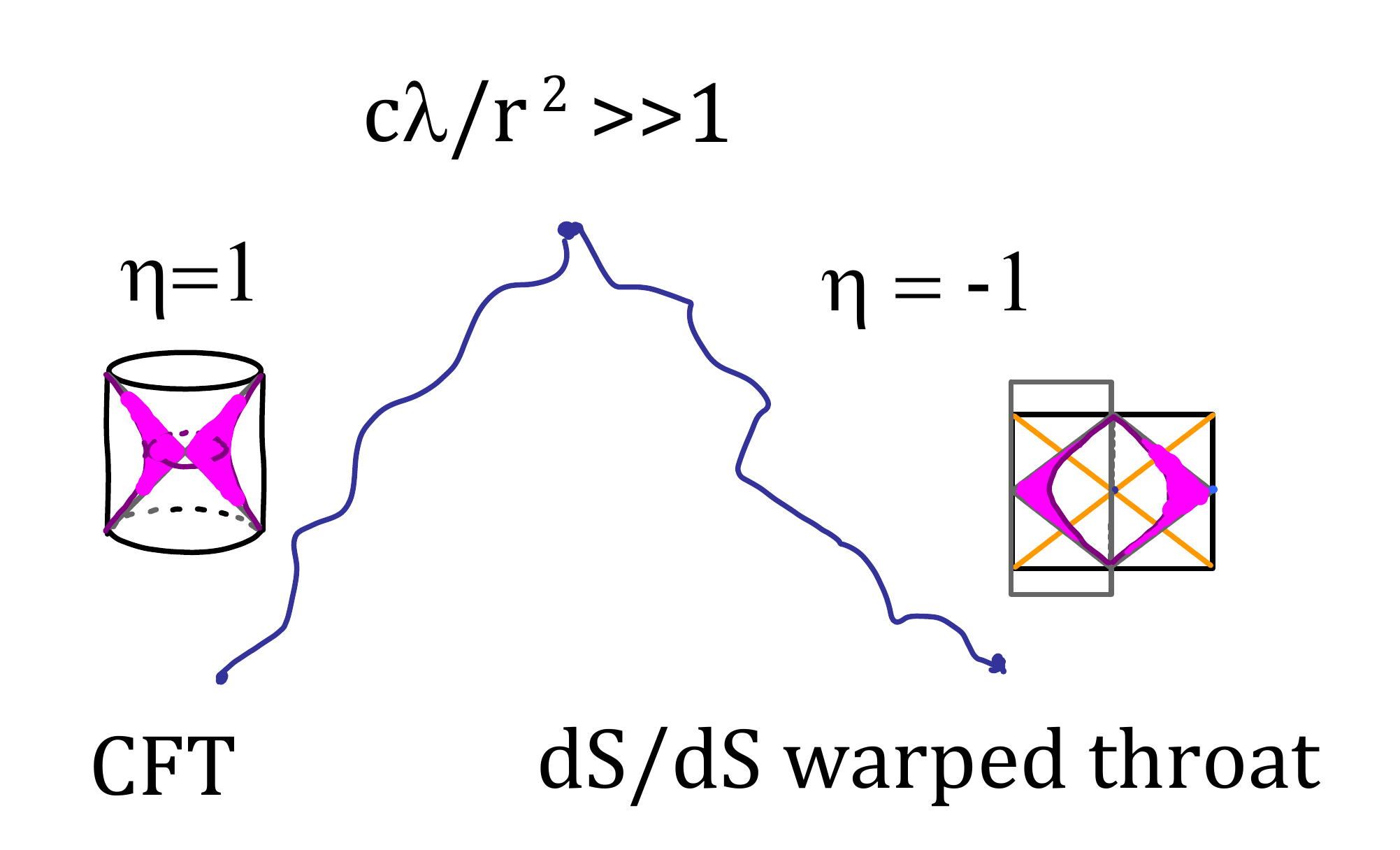}
\end{center}
\caption{The reconstruction of the $dS/dS$ throat from a seed CFT proceeds via two joined trajectories as described in the text.  The first trajectory (on the left) evolves the system from a pure CFT, via a sequence of cutoff AdS/dS systems, to the limit where this cutoff scale goes to zero, indicated by the point at the top of the figure.  That point is the start of a new trajectory incorporating $\Lambda_2\propto \eta-1$, with increasing cutoff scale, culminating in the full $dS/dS$ warped throat. }
\label{TwoTrajectories}
\end{figure}

Let us now switch to the $dS_3$ case. Clearly the deformation should be modified in some way. As we already anticipated, there is another natural parameter in the $2d$ theory -- the cosmological constant $\Lambda_2$.
This term can enter nontrivially into the trajectory because it contributes to the stress-energy tensor, which is recomputed at each step as we evolve along the trajectory.  
Let us study its effect to see if this is enough to produce the new contribution we derived on the gravity side in \S\ref{sec:bulk}, in particular the last term in (\ref{flowQFTvar}) with $\eta=1$ for the $AdS_3$ case and $\eta=-1$ for the $dS_3$ case.   We need a one-dimensional trajectory in the space of $2d$ theories, so dimensional analysis implies that $\Lambda_2$ should be proportional to $1/\lambda$. In our conventions, as we will see shortly, the needed dimensionless coefficient is such that under the infinitesimal changes in $\lambda$
\be
d\Lambda_2=\frac{(1-\eta)d \lambda}{2\pi \lambda^2}\ = \frac{d \lambda}{\pi \lambda^2}\;.
\ee  
Then \eqref{eq:ZTTbnew} gets modified in the following way:
\be\label{eq:ZTTbLambda}
\frac{\partial}{\partial \lambda}\, \log Z =-2\pi \,\int d^2 x\,\sqrt{g}\, \langle T \bar T \rangle+\frac{1}{\pi \lambda^2}\int d^2 x \,\sqrt{g}=-8\pi^2r^2\,\langle T\bar T \rangle\,+\frac{4 r^2}{\lambda^2}\,.
\ee
Again differentiating with respect to $r$ leads to 
\be
\label{eq:SZdS}
4\, \frac{\d}{\d\lambda}\, E(r,\lambda) = E(r,\lambda)\, \frac{\d}{\d r} \,E (r,\lambda)+\frac{8 r }{\lambda^2}\,.
\ee
It is immediate to check that now \eqref{eq:alphadS} is a solution. The above equations can be easily generalized to the case of tall $dS_2$, reproducing \eqref{Ttautauanswertext} for any $\mu$.   Below and in Fig. \ref{TwoTrajectories}\ we will specify the boundary conditions needed for our reconstruction of the $dS_3/dS_{2\mu}$ warped throat from a seed CFT.  

We conclude that for our symmetric states the $(A)dS_3$ quasi-local stress tensor is reproduced by the following trajectory in the space of large-N $2d$ theories:
\be\label{DeltaLlambda}
\delta {\cal L} =  \delta\lambda\left\{ 2\pi T\bar T-\frac{1-\eta}{2\pi\lambda^2}\right\}\,.
\ee
where ${\cal L}$ is the Lagrangian.  Schematically, the action $S=\int \cal L$ is related to $W=-i\log(Z)$ by an integral transform
\be\label{WS}
e^{i W[J]}=\int DM e^{i S[M] + i \int J {\cal O}}
\ee
in terms of the matter $M$, sources $J$ and operators ${\cal O}$ of the theory.  
Let us take ${\cal O}$ to be the stress-energy tensor, whose source is the change in the boundary metric $\delta g_{\mu\nu}$.   Considering a semiclassical large-$c$ saddle point, we have 
\be\label{semiW}
W \simeq S_\text{saddle}+\int \delta g_{\mu\nu} T^{\mu\nu}_\text{saddle} ~~~~ \text{large}~c
\ee
in terms of the action evaluated at the saddle (which could be written in terms of the appropriate `master fields', captured by the classical physics on the gravity side in a holographic theory \cite{Maldacena:1997re}).   Each gravity-side geometry that we wish to reproduce, at a given point along the trajectory of theories, corresponds to such a large-$c$ saddle point, with both the operator and the source taking definite values.   When we define our trajectory in the space of $2d$ theories to reproduce these configurations, we could in principle do so either in terms of either $W[g_{\mu\nu}]$ or $S[T_{\mu\nu}]=\int {\cal L}$.  We expressed this in terms of the action in (\ref{DeltaLlambda}).  

Since bulk energy levels, and hence energy levels of a deformed CFT, satisfy the trace-flow equation (\ref{flowQFTvar}) have derived this equation for these configurations from the deformation of the action. Previously it was derived on the $3d$ gravity side. Note that from the $2d$ point of view (\ref{flowQFTvar}) includes the conformal anomaly and it was not a priori obvious that the anomaly contribution would automatically remain the same all along a trajectory that  deviates from a pure CFT.\footnote{See \cite{Hartman:2018tkw}\ for some comments on this point as well as the relation between $W$ and the effective action.}  Instead it could have been necessary to tune the $2d$ trajectory in order to maintain this, via a Wess-Zumino term. We just showed that, at least in large-$N$ theories on homogeneous states, this relation indeed holds.\footnote{In this regard it is interesting to note a difference between this $d+1=3/d=2$ duality and the higher-dimensional generalization to $d+1=4, d=3$, the case of phenomenological interest. Following \cite{Hartman:2018tkw}\ we would not have the anomaly, but would have an extra contribution to ${\rm Tr}\,T$ with a shift  in $T_{\mu\nu}$ proportional to the Einstein tensor of the $3d$ boundary.  For our $dS_3$ boundary, this in turn is proportional to the metric, and would contribute a term similar to the one parameterized by $b_{CT}$ above in (\ref{bCTT}). 
}

At large $c$, we may consider less symmetric excitations which should match those of the quasilocal stress energy tensor computed via general relativity on the gravity side.
For non-symmetric states  we may follow essentially the same steps.  Instead of the relation between $T^\phi_\phi$ and  $T^\tau_\tau$ enforced by symmetry, we can now use the ``pressure'' relation, 
\be\label{eq:pressure}
\langle T^\phi_\phi \rangle= \frac{d (L \langle T^\tau_\tau \rangle)}{dL}\,.
\ee 
The variation with respect to the global $dS_2$ circle size $L$ is independent of the curvature ${\cal R}^{(2)}$ since we may change the periodicity of the $\phi$ direction independently of the curvature.\footnote{We saw this above in our family of tall de Sitter solutions parameterized by the $\phi$ period $2\pi\mu$.}  The differential equation for the energies will now become a partial differential equation in three variables, which is harder to manage. It may be more straightforward to use the trace-flow relation  (\ref{TphiTtautext}) in this case, which we have not proven in general, but which is now very plausible since the anomaly contribution is likely independent of the details of the state. Applying it along with the preasure relation gives a differential equation
\be\label{diffL}
\langle T^\phi_\phi \rangle = \frac{d (L\,\langle T^\tau_\tau \rangle)}{dL} = \frac{\langle T^\tau_\tau \rangle+c \frac{{{\cal R}^{(2)}}}{24\pi} +\frac{\eta-1}{\pi\lambda}}{\pi\lambda \langle T^\tau_\tau \rangle-1}\,,
\ee
which has solutions
\be\label{dressedLsolns}
\langle T^\tau_\tau \rangle=\frac{1}{\pi\lambda}\left(1-\sqrt{\eta+c \frac{{{\cal R}^{(2)}\lambda}}{24}- \frac{C_1\lambda}{L^2}} \right)\,.
\ee
We note that this equation admits more general solutions than the vacuum configurations we obtained as the ground state in each tall dS sector above (\ref{Ttautauanswertext}); those followed from a second algebraic equation $ \langle T_{\mu\nu} \rangle=\alpha g_{\mu\nu}$ which does not apply in general.   It will be an interesting future direction to analyze in detail the bulk dual of more general states than the vacuum states in each sector, along the lines of \S\ref{generalconjecture}.   

Finally let us give the boundary conditions on our trajectories, to complete our derivation of the bulk geometry.  We do this using the following procedure -- see Fig. \ref{TwoTrajectories}.  
We start from a seed CFT, which we will describe in terms of a Lagrangian for simplicity.   We then turn on $T\bar T$ and follow the trajectory \cite{Donnelly:2018bef} dual to $AdS_3$ cut off on a $dS_2$ (or tall $dS_2$) slice.   This AdS/dS trajectory satisfies the trace flow equation with $\eta=1$.  At the end of this trajectory, $c\lambda/r^2\to \infty$, the corresponding radial cutoff on the gravity side is deep in the infrared region ($w_c\to 0$), where it is indistinguishable from a dS/dS throat with a vanishingly small cutoff scale.   Below in \S\ref{sec:epsilon} we will comment further on the physics of this region with $r<\ell$.   We take that as the seed for the new trajectory which we formulated above in (\ref{DeltaLlambda}), which generates the trace flow equation with $\eta=-1$.  The new ingredient we introduced is an incremental shift of the $2d$ cosmological constant $\Lambda_2$ at each step in the trajectory.  A shift of the vacuum energy by itself would not have an effect in quantum field theory, but it enters nontrivially in the $T\bar T$ trajectory:  $\Lambda_2$ contributes to the stress-energy tensor which must be recalculated at each step, appearing nonlinearly in the $T\bar T$ contribution to the trajectory.  Following that trajectory step by step gives a dual construction of a sequence of radially cut off $dS_3/dS_2$ warped throats, ultimately reaching its endpoint at $w_c=\ell\,\pi/2$.  {We would like to point out some similarities between this construction and the construction of \cite{Giveon:2017nie}. There it was shown that a single-trace $T\bar T$-like deformation of the boundary CFT produces the spacetime that interpolates between $AdS_3$ in the IR and a linear dilaton spacetime in the UV.}

Let us finally comment on the sphere partition function of the $2d$ theory. Once we know the stress tensor it is straightforward to integrate \eqref{eq:Trace} to obtain it as in \cite{Donnelly:2018bef}. One subtlety, however, is that integration introduces an integration constant which is related to the cutoff dependence of the sphere partition function. In AdS we may fix this dependence by matching in the $r^2/\lambda\to\infty$ limit of the CFT partition function, 
\be\label{eq:ZCFT}
\log Z_\text{CFT}(r) = \frac{c}{3}\, \log\,\frac{r}{\epsilon}+\ldots\,.
\ee
The result is then
\be\label{eq:ZAdS}
\log Z_{\text{AdS}}(r) = - \frac{4r^2}{\lambda}\,\left(1-\sqrt{\frac{\lambda c}{12 r^2}+1} \,\right)+\frac{c}{3}\,\log \left[\frac{r}{2\epsilon}\left(1+ \sqrt{\frac{\lambda c}{12 r^2}+1}\,\right) \right]\,.
\ee
We note that this prescription and the cutoff dependence should be taken with care for a number of reasons.   First, there is an energy cutoff naturally present in the $T\bar T$ deformed theories for the sign of the coupling we consider, and there is a radial cutoff on the gravity side in our construction of the quasilocal stress tensor.  See section~\ref{sec:epsilon} for more discussion of these issues.   

At a technical level, our expression (\ref{eq:ZAdS}) differs from Eq. (2.4) of \cite{Donnelly:2018bef}, which implicitly chose $\epsilon \sim \sqrt{\lambda c}\sim \ell$ from the start via a boundary condition at $\lambda/r^2=\infty$. With $\epsilon$ tied to $\lambda$ from the start, the partition function would not satisfy the $T\bar T$ equation \eqref{eq:ZTTbnew} due to the extra contributions to the derivative with respect to $\lambda$.  Instead, we keep $\epsilon$ independent of $\lambda$ in deriving the differential equation above; it can later be fixed to a particular value.  

In any case, in this work we focus on $\epsilon$-independent quantities.  One motivation for that is the evidence found in \cite{Dubovsky:2012wk, Dubovsky:2017cnj, Dubovsky:2018bmo} that the $T\bar T$-deformed theory is not a local quantum field theory, so the corresponding divergences may not enter into its observables in the usual way.

For the dS case, and with the integration constant determined by matching to \eqref{eq:ZAdS} at $r^2/\lambda\to 0$, we get
\be\label{eq:ZdS}
\log Z_\text{dS}(r) = \frac{c}{3}\,\log\,\frac{\sqrt{\lambda c/12}}{2\epsilon}- \frac{4r^2}{\lambda}\,\left(1-\sqrt{\frac{\lambda c}{12 r^2}-1} \,\right)+\frac{c}{3}\,\arctan \left(\frac{1}{\sqrt{\frac{\lambda c}{12 r^2}-1}} \right)\,.
\ee

\subsection{Relations among parameters and large-$c$ physics}\label{subsec:parameters}

Having defined the $2d$ trajectory, let us now collect our dictionary for parameters and relations between them.  
Although we initially kept track of $b_{CT}$ on the gravity side, we will set it to 1 here as in the previous section.
On the gravity side, we then have four continuous parameters: $G, \ell=\ell_{(A)dS_3}, w_c, \mu$.  We also have the sign $-\eta$ of the bulk cosmological constant.

On the QFT side, we have $c, \Lambda_2, \lambda$, the curvature radius $r$, and the size of the circle $L$ at $\tau=0$.  There is one more parameter here, but as we discussed in the previous subsection our flow is obtained by maintaining one relation between them at each step, $\Lambda_2\propto 1/\lambda$.  

The dictionary between the two sides and the one relation is
\bea\label{relations}
c &=& \frac{3\ell}{2G} \nonumber\\
\lambda &=& 8 G\ell \nonumber\\
r&=&\ell \sin(\frac{w_c}{\ell}) \nonumber\\
L &=&2\pi \mu \ell \sin(\frac{w_c}{\ell}) \nonumber\\
\Rightarrow ~~ \Lambda_2=-{\frac{1}{\pi\lambda}}&=& \frac{c}{12}\frac{1}{r^2} ~~~~ (w_c=\pi/2)
\eea
where the last relation was applied for $w_c$ at the UV slice $w_c\to\pi/2$.  

We note that (for general $w_c$) $\lambda$ is of order $1/c$ while $\Lambda_2$ is of order $c$.   This is why the relevant deformation within our flow and its interplay with the $T\bar T$ deformation is significant at the level of our large-$c$ analysis.   In the bulk AdS analysis of \cite{McGough:2016lol, Donnelly:2018bef, Kraus:2018xrn}, the leading effect of the $\lambda T\bar T$ deformation was to cut off the throat, leaving the bulk $AdS_3$ geometry intact up to $1/c$ corrections.   In that context, the $T\bar T$ deformation plays a leading role for the energy levels since BTZ black holes are heavy objects with energy of order $c$.  In our case, even for the ground state at $m=0$, the throat is modified compared to the AdS case, and this is reflected in the order-$c$ contribution $\Lambda_2$.  In the configurations we have studied, all terms in the trace flow equation \eqref{flowQFTvar}\ are of order $c$.

\subsection{Comments on energy scales and the regime $r<\ell$}\label{sec:epsilon}

The boundary condition for the part of our trajectory that builds up the $dS/dS$ throat (summarized in Figure \ref{TwoTrajectories}) occurs in the regime $\frac{r^2}{c\lambda}\sim \frac{r^2}{\ell^2}\ll 1$ where $r$ is the curvature radius of our $2d$ spacetime, equal to the radial position of the boundary on the gravity side.  Here we comment on the physics contained in the system as a function of $r/\ell$, including the small-radius regime.  

In the AdS/CFT dictionary, one expects that $c$ degrees of freedom are required to fully reconstruct the physics in a region of size $\ell$ \cite{Susskind:1998dq}, and capturing the physics below this scale may require fine details of the dual field theory.  In the bulk $dS_3$ case, with the simple particle states that introduce a deficit angle, this may benefit from generalizing the recent progress \cite{Balasubramanian:2014sra, Balasubramanian:2016xho, Balasubramanian:2018ajb}\ for particle states in $AdS_3$.   
In this work we have focused on a limited class of  low-lying stress-energy configurations, described in terms of hydrodynamic variables, which are not sensitive to all of the microphysical details.    
In this section we review more generally the energy scales of various states that fit inside a radial cutoff at $r$ (related
to a cutoff on total energy entering into the recent entropy calculations \cite{Dong:2018cuv}).

Starting with the bulk AdS case, we have BTZ black holes for $r>\ell$.  These have mass $M_{BTZ}\sim \frac{r_{BH}^2}{G\ell^2}$ in terms of the horizon radius $r_{BH}$, such that with our radial cutoff these are bounded by $M_{BTZ} < \frac{r^2}{G\ell^2}$.  We can write this as a bound
\be\label{BTZcutoff}
E_{BTZ} < \frac{c}{\ell}\left(\frac{r}{\ell} \right)^2, ~~~~~~ r>\ell.  
\ee
In both the bulk AdS and bulk dS case, for $r<\ell$ we have no black hole horizons, but we do have particle states which source a deficit angle as in (\ref{mumass}).  In this regime, the energy is bounded by $1/G$, 
\be\label{mcutoff}
E_{particle}\le \frac{c}{\ell}, ~~~~~~ r<\ell
\ee
In both regimes, there are many types of excitations that fit below this cutoff.  These are perhaps analogous to near-horizon excitations studied in the context of holographic hydrodynamics \cite{Rangamani:2009xk}.  
For example a test particle in $(A)dS_3$ of proper energy $E_{pr}\ll 1/G$ has redshifted energy
\be\label{Ered}
E=\sqrt{-g_{00}}E_{pr} < \frac{1}{\ell} (r E_{pr})
\ee
where the inequality in the last expression represents the radial cutoff.  For a particle to fit in a region of size smaller than $r$, it needs proper energy at least of order $1/r$.  The configurations we have focused on in this work fit below the cutoff:  they are homogeneous, vacuum configurations of the stress-energy in each sector parameterized by $\mu$ (\ref{dS2tall}).  

Let us finally compare this energy cutoff to the scale of the dynamical cutoff on bare energies that arises from the dressing by $T\bar T$.   Working with the original cylinder case,  from (\ref{TFEpressureTalk}), we have 
\be\label{Tbound}
E^{(0)} < \frac{L}{2\pi\lambda} \sim \frac{c}{\ell}\cdot\frac{r}{\ell}
\ee
For any value of $r/\ell$, this is below the cutoff (\ref{BTZcutoff}) or (\ref{mcutoff}).

\subsection{Comment on more general configurations}\label{generalconjecture}

We have engineered a trajectory in the space of $2d$ theories which produces the cutoff $dS_2$ and tall $dS_{2\mu}$ configurations that we studied on the gravity side, realizing our initial goal of reconstructing the bulk de Sitter geometry via a well-defined holographic dual.    

As we noted above, the solutions (\ref{dressedLsolns}) to our differential equation for $\langle T^\tau_\tau \rangle$ go beyond the vacuum solutions  (\ref{Ttautauanswertext}) in each $\mu$ sector.  The system can be excited in a much wider variety of ways, well beyond the homogenous configurations treated in (\ref{Ttautauanswertext}) and (\ref{dressedLsolns}).   Let us briefly formulate a conjecture for these and their dual descriptions.  

On the gravity side, we can introduce a boundary which is fixed by a Dirichlet condition to be $dS_{2\mu}$ for some $\mu$, and excite the system with some configuration of sources.  In some cases it may be natural to choose $\mu$ according to the rest mass of a collection of sources, but in any case it can be fixed.    Having fixed this boundary condition and regularity in the interior, one can in principle determine the solution to the bulk field equations and calculate the quasilocal stress-energy tensor.        
The conjecture is that the dual $2d$ theory we formulated in this section lives on $dS_{2\mu}$, and contains excited states whose stress-energy tensor is given by the quasilocal stress-energy tensor of the gravity side.

\section{Entanglement entropy}\label{sec:EE}

The entanglement entropy (EE) provides an interesting probe of the dynamics, with nontrivial implications in both sides of the duality. In this section we will compute the EE in the presence of the field theoretic trajectories studied in the preceding sections.   The recent work \cite{Dong:2018cuv} calculated the entanglement and Renyi entropies of the density matrix for one of the two matter sectors in the dS/dS correspondence, finding a flat entanglement spectrum at large $c$.  In the present work, we are formulating each matter sector separately (as a cut off warped throat with positive bulk cosmological constant).  

In the semiclassical regime of the bulk gravity side, the EE is captured by the Ryu-Takayanagi formula \cite{Ryu:2006bv, Hubeny:2007xt}. On the other hand, in general it is very hard to compute the EE directly in the field theory dual. However, it was found in \cite{Donnelly:2018bef} that the two-dimensional calculation can be carried out explicitly when one considers antipodal points in the $dS_2$ spacetime. We will then restrict to this case, and we will compare the results from both sides of the duality, finding a nontrivial agreement.

\subsection{Holographic entanglement entropy}

We choose a surface at time $\tau=0$, for which the gravity side metric reads
\be
ds^2= dw^2 + \ell^2\,\sin(h)^2\frac{w}{\ell}\,d\phi^2\,.
\ee
The circle runs between $\phi \in (-\pi, \pi)$, and we wish to calculate the EE for an interval specified by antipodal points in this circle, e.g. $\phi \in (-\pi/2,\pi/2)$, working in the Euclidean vacuum. The Dirichlet wall is taken at $w=w_c$, and the corresponding radius of the circle is
\be\label{eq:circler}
r= \ell \sin(h)\frac{w_c}{\ell}\,.
\ee

The minimal surface is described by some curve $\phi =  \phi(w)$ anchored at $w_c$, obtained by extremizing the area
\be
A =2 \int_{w_t}^{w_c} dw\,\sqrt{1+ \ell^2 \sin(h)^2\frac{w}{\ell} \left( \frac{d\phi}{dw}\right)^2}\,.
\ee
Here $w_t$ is the turning point of the curve, where $d\phi/dw \to \infty$. For the special case of antipodal points, it is not hard to see that the solution is $w_t=0, d\phi/dw=0$. By symmetry, the solution is simply a straight line along $w$, connecting the two end-points of the interval. This is similar to what happens in the Rindler case (a semi-infinite interval).

The result for the EE is then
\be
{\cal S}= \frac{A}{4 G}=\frac{w_c}{2G}\,.
\ee
From (\ref{eq:circler}), this can be rewritten in terms of the circle radius:
\be
{\cal S}(r) =\frac{l}{2G}\,\arcsin(h)\,\frac{r}{\ell}\,.
\ee
For comparison with the $2d$ dual, let us write this in terms of field theory parameters,
\be\label{eq:Sboth}
{\cal S}(r) =\frac{c}{3}\,\arcsin(h)\,\frac{r}{\sqrt{\lambda c/12}}\,.
\ee

\subsection{Entanglement entropy in the $dS_2$ dual:  DS/dS/dS}

Let us now focus on the EE for the $2d$ theory on $dS_2$ at $\tau=0$, and with an interval defined by antipodal points of the de Sitter circle. The radius of the circle is denoted by $r$. The key simplification in this case, noted in \cite{Donnelly:2018bef}, is that there is a geometric symmetry that enables us to perform the replica trick for arbitrary non-integer $n$.  In this section we apply the method \cite{Donnelly:2018bef}\ to compute a derivative of the entanglement entropy, generalizing to the $dS/dS$ case.   
      
The partition function on the branched cover is given by\footnote{We thank H. Casini for discussions on these aspects.}
\be
\frac{d}{dn}\,\log Z_n \Big|_{n=1}=- \int d^2x \,\sqrt{g}\,\langle T^{\tilde\phi}_{\tilde\phi} \rangle\,.
\ee
where $\tilde\phi$ is the Euclidean rotation of the static time coordinate.  Here we consider a static observer on a worldline in the center of our interval, i.e. at $\phi=0$.   Their thermal entropy in the Euclidean vacuum corresponds to tracing out the region outside their horizon, which is the region outside of the interval.         

The EE then becomes
\be\label{eq:Sn}
{\cal S}(r) =\left(1- n\frac{d}{dn}\right)\log Z_n \Big|_{n=1}=\left(1- \frac{r}{2}\frac{d}{dr}\right)\log Z(r)\,,
\ee
where in the final step we used $\langle T^{\tilde\phi}_{\tilde\phi} \rangle =\frac{1}{2}\langle {\rm Tr}\, T \rangle$ for the dS vacuum, together with (\ref{eq:Trace}).

Since the entanglement entropy is divergent in relativistic quantum field theory,  and there are indications that the $T\bar T$ deformed theory does not have arbitrary local observables \cite{Dubovsky:2012wk, Dubovsky:2017cnj}, we will focus on the cutoff-independent quantity
\be\label{eq:Cfc}
C(r) = r \frac{d {\cal S}}{dr}\,,
\ee
which plays the role of a running C-function in RG flows \cite{Casini:2012ei}.  
Using (\ref{eq:Sn}) and
\be
r \frac{d \log Z}{dr}= -\int d^2 x \,\sqrt{g}\, \langle T^i_i \rangle=-\frac{8 r^2}{\lambda}\,\left(1-\sqrt{\frac{\lambda c}{12 r^2}+\eta}\, \right)\,,
\ee
(where we used the ground-state expectation value 
$\langle T_{ij} \rangle = \alpha g_{ij}$  
with $\alpha$ given in \S \ref{subsec:ground-state}), yields
\be\label{eq:Cfc2}
C(r) = \frac{c}{3}\,\left(\frac{\lambda c}{12 r^2}+\eta\right)^{-1/2}\,.
\ee
This matches precisely the running C-function computed from the gravity result (\ref{eq:Sboth}). The EE then provides a nontrivial consistency check of the conjectured duality in the presence of the $T \bar T$ and $\Lambda_2$ flows.

\begin{figure}[htbp]
\begin{center}
\includegraphics[width=10cm]{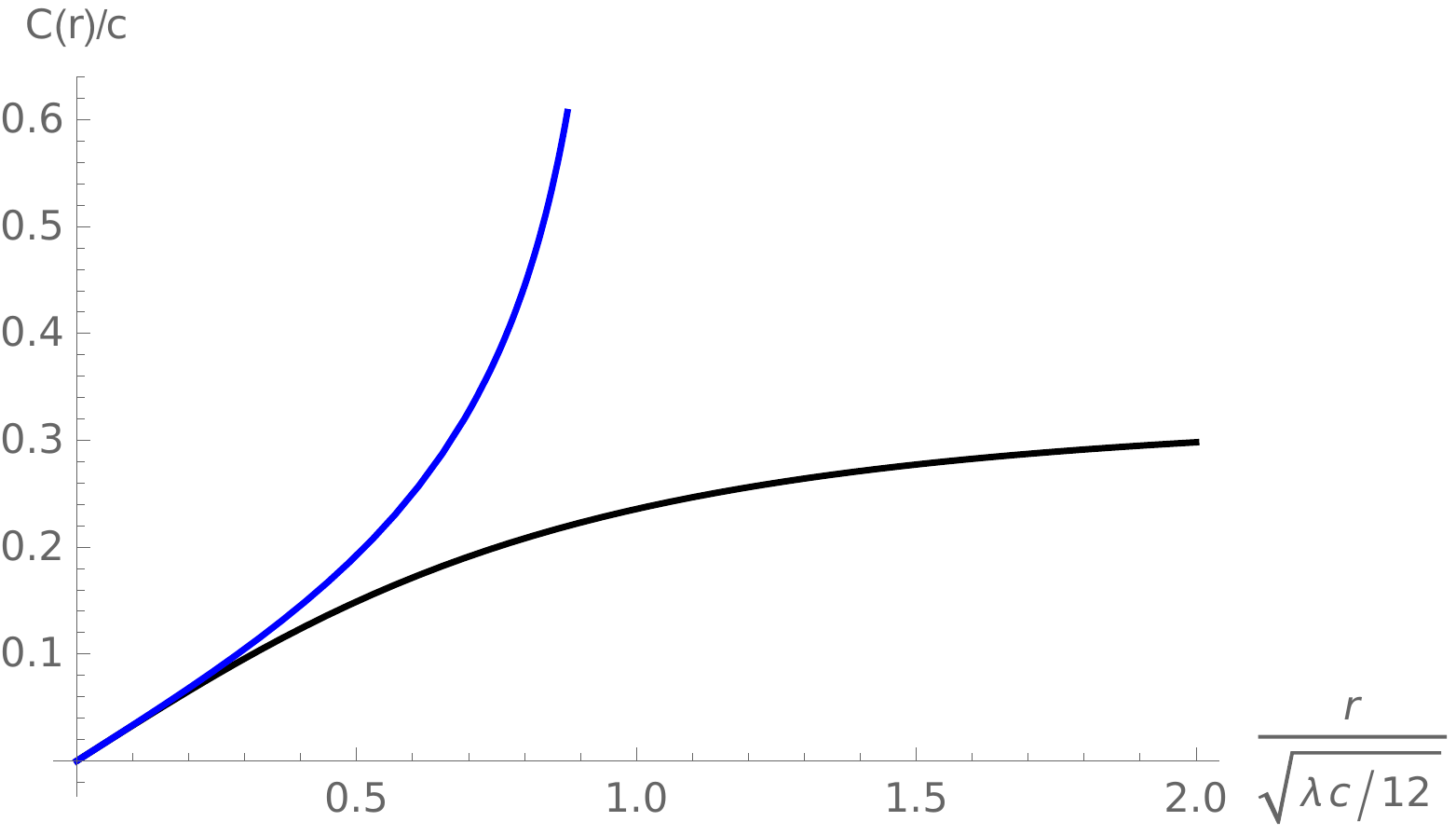}
\end{center}
\caption{Function $C(r)/c$ for different $dS_2$ radii $r$. The black line corresponds to $AdS_3$, and the blue line is for $dS_3$. In the second case, $C(r)$ diverges at $r=\sqrt{\lambda c/12}$.}
\label{fig:EE}
\end{figure}

It is interesting to compare the AdS and dS results in order to understand the effect of $\Lambda_2$. This is shown in Fig. \ref{fig:EE} for different values of the $dS_2$ radius. Both results are approximately the same for small $r$; this is a consequence of our matching procedure in the IR region. On the other hand, as $r$ increases, $\Lambda_2$ leads to an increase in the C-function compared to the $AdS_3$ result. In particular, $C(r)$ diverges at the largest allowed radius $r=\sqrt{\lambda c/12}=\ell$, the UV slice of $dS_3$. Instead, as expected, in the $AdS_3$ dual $C(r) \to c/3$ as $r^2/\lambda \to  \infty$. 

We should stress that the entanglement entropy analyzed here can differ in qualitative ways from the entanglement for a fixed interval in Minkowski space. This is because we have put the theory on a curved space, and also finite size effects are largest for our choice of interval size. In particular, we note that the running C-function does not satisfy the strong subadditive inequality $C'(r) \le 0$ that holds in Minkowski space \cite{Casini:2012ei}. It would be interesting to try to compute the EE on a fixed interval in Minkowski space, for a CFT deformed by $T \bar T$ and/or $\Lambda_2$.

\section{Conclusions and Questions}\label{sec:conl}

In this work, we derived a trajectory in the space of quantum field theories which reproduces the gravity-side geometry of the dS/dS correspondence, including a sequence of massive particle excitations and the entanglement entropy.  We worked at large $c$ and focused on the universal gravitational sector, but expect similar methods to apply to incorporate additional bulk matter \cite{Taylor:2018xcy, Hartman:2018tkw, Baggio:2018rpv, Chang:2018dge, usnext}.  Starting from any seed CFT, our trajectory combining $T\bar T$ and $\Lambda_2$ deformations produces a stress-energy tensor that matches the quasilocal stress-energy of the bulk $dS_3$  theory.   Although the $T\bar T$ coefficient is of order $1/c$, and left the infrared throat intact in the AdS/CFT version \cite{McGough:2016lol}, the $\Lambda_2$ deformation is of order $c$, leading to an order one effect on the throat geometry even below the cutoff scale in our case.  Our bulk dS case also has favorable properties such as causal propagation of perturbations around back-reacted particle states, simply because these do not have black hole horizons.   

It will be interesting to consider the effect of bulk matter fields on the dual trajectory, a point stressed in \cite{Kraus:2018xrn}\ in the bulk AdS case; see also \cite{Hartman:2018tkw, Taylor:2018xcy}.  The agreement of the dressed energies at the level of pure gravity in the bulk, now in dS as well as AdS, seems remarkable in itself.  The dS case provides additional motivation for formulating the bulk matter field generalization in both cases.  In this regard, we recall the earlier holographic RG study in \cite{Dong:2012afa}\ which showed that the bulk dS and bulk AdS cases share an important property:  single-trace operators do not flow, and higher-trace operators are determined in terms of lower-trace ones.   The $T\bar T$ deformation and our generalization here do not constitute an ordinary RG flow, but rather they specify a trajectory in the space of couplings.  So far, both the holographic RG including bulk matter and gravity, as well as the pure gravity sector of the generalized $T\bar T$ deformation, readily generalize from the bulk AdS to bulk dS cases.  We suspect that the same will be true for the generalization of $T\bar T$ needed to capture bulk matter fields; in any case this will be very interesting to investigate.  

A related question is the role of string theory, even at the classical level, for which target space boundaries are not arbitrary.  In this regard the string-theoretic treatment of the multiple trace deformations introduced in \cite{Aharony:2001pa}\ may play a role in the deformation of our seed CFT, since a Dirichlet condition on a finite radial slice may be related to a non-local boundary condition at the AdS boundary \cite{Brattan:2011my, Marolf:2012dr}.\footnote{We thank M. Rangamani and J. Maldacena for discussions of this.}

This raises numerous additional questions for further research. { In \cite{Dubovsky:2017cnj,Dubovsky:2018bmo}, the pure $T\bar T$ deformation  on 2d spacetimes with zero curvature was shown to match a version of Jackiw-Teitelboim gravity in detail; other interesting works relating $T\bar T$ to metric fluctuations appeared in \cite{Cardy:2018sdv, Cottrell:2018skz}}.  It would be very interesting to understand if there is a similar statement in our case.   In our analysis, we found it most straightforward to work with the $2d$ theory on the `tall de Sitter' spacetime (\ref{dS2tall}).   Such a spacetime arises in the presence of excitations of de Sitter via the constraints of gravity, ultimately closing up its neck if too much energy is introduced.  We have not investigated the connection to $2d$ gravity in this paper, but it may line up with the simple role that tall dS played in our analysis here.

The $T\bar T$ deformation at large $c$ may also apply to other ways of organizing de Sitter holography \cite{Anninos:2012qw, Anninos:2017hhn}.  Within a static patch, the simplest slicing of the geometry is by cylinders, and it would be interesting to consider cutting that spacetime off at different radial positions, computing the quasilocal stress tensor, and engineering a trajectory in the space of $2d$ theories on flat spacetime which reproduces it.     

Finite $c$ will be important to understand.  It is at finite $c$ that some of the basic effects relevant to de Sitter physics such as quantization of fluxes and metastability arise.  The original analysis of the $T\bar T$ deformation in \cite{Zamolodchikov:2004ce, Smirnov:2016lqw, Cavaglia:2016oda} did not require or involve large $c$, but it only applies directly for flat 2d spacetime.   It will be interesting to see how far the isometries of $dS_2$ go toward factorization of $T\bar T$ correlators.  On the other hand, complications that may arise at finite $c$ could relate to the ultimate metastability of de Sitter.   In some subset of cases, in fact, the de Sitter we build up could have perturbative instabilities, perhaps related to relevant deformations of the seed CFT.  This will also be interesting to understand from a model-building perspective.  

In this work we focused on a single warped throat (among the pair of them in dS/dS (\ref{dSdSmetric})), generalizing the cutoff AdS throat of \cite{McGough:2016lol}.  In particular, this involved a single CFT among the pair of identical ones that enter in the full dual of dS.  It would be very interesting to analyze the { overall} $T\bar T$ deformation, perhaps combined with relevant directions as we did here, of the pair of identical matter sectors in the dS/dS correspondence.   That by itself introduces interactions between the two sectors, a key feature of the combined system \cite{Alishahiha:2004md, Alishahiha:2005dj, Dong:2018cuv}.  
In any case, in this work we have seen how to produce the warped throat building blocks of the dS/dS correspondence, concretely on the field theory side, via a generalization of the $T\bar T$ deformation.

\vskip 5mm

\noindent{\bf Acknowledgements}

\smallskip

\noindent We are grateful for useful discussions with Jerem\'ias Aguilera Damia, Vijay Balasubramanian, Horacio Casini, John Cardy, Billy Cottrell, Xi Dong, Sergei Dubovsky, Dan Freedman, Jorrit Kruthoff, Juan Maldacena, Mehrdad Mirbabayi,  XiaoLiang Qi, and Mukund Rangamani.
The research of ES is supported in part by the  National Science Foundation under grant number PHY-1720397, a Simons Foundation Investigator Award, and VG and ES are supported by the Simons Foundation Origins of the Universe program (Modern Inflationary Cosmology collaboration). VG is the Della Pietra Postdoctoral Member at IAS.  GT is supported by CONICET (PIP grant 11220150100299), ANPCYT PICT grant 2015-1224, UNCuyo and CNEA.

\appendix

\section{GR side relations}\label{app:GReqs}

In this Appendix we collect some useful formulas from gravity-side GR.

\subsection{Stress tensor flow equation}\label{subsec:appflow}

We work in Lorentzian $(-++)$ signature, and gauge-fix the metric to the form
\be
ds^2 = dw^2 +g_{ij}(w,x)\,dx^i dx^j\,.
\ee
The bulk Einstein equations for the action (\ref{action}) are
\be
E^a_{\;b}= R^a_{\;b}-\frac{1}{2}\delta^a_b\,R-\delta^a_b\,\frac{\eta}{\ell^2}=0\,.
\ee
In components,
\bea
E^w_w&=& \frac{1}{2}(K^2-K^{ij}K_{ij})-\frac{1}{2} {\mathcal R}^{(2)}-\frac{\eta}{\ell^2}=0\nonumber\\
E^w_j&=& \nabla^i(K_{ij}- g_{ij}K)=0 \\
E^i_j&=&-\partial_w(K^i_j-\delta^i_j K)-K K^i_j+\frac{1}{2}(K^{mn}K_{mn}+K^2)-\delta^i_j\,\frac{\eta}{\ell^2}=0\,, \nonumber
\eea
where $K_{ij}=\frac{1}{2} \partial_w g_{ij}$ is the extrinsic curvature at fixed $w$, and ${\mathcal R}^{(2)}$ is the Ricci scalar of $g_{ij}$.

As in \cite{Kraus:2018xrn}, let us rewrite the constraint equation $E^w_w=0$ in terms of the quasi-local stress tensor (\ref{bCTT}). We have
\be
T^{ij} T_{ij}-(T^i_i)^2 = \frac{1}{(8 \pi G)^2} \left(K^{ij} K_{ij}- K^2+2\frac{b_{CT}}{\ell} K-2 \left(\frac{b_{CT}}{\ell} \right)^2\right)\,,
\ee
and from the trace of (\ref{bCTT}),
\be
K = -8 \pi G T^i_i+2\frac{b_{CT}}{\ell}\,.
\ee
Replacing these relations into $E^w_w=0$ yields
\be
T^i_i =-\frac{4\pi G \ell}{b_{CT}} (T^{ij} T_{ij}-(T^i_i)^2)-\frac{\ell}{16 \pi G\, b_{CT}} {\mathcal R}^{(2)}-\frac{\eta-b_{CT}^2}{8\pi G \ell\, b_{CT}}\,.
\ee
In particular, the solution for the ground state is
\be\label{alpha2}
\alpha = \frac{1}{8\pi G \ell}\left(b_{CT} - \sqrt{\frac{\ell^2}{r^2}+\eta} \,\right)\,.
\ee

Setting $b_{CT}=1$ we arrive to (\ref{TequationdS}). We interpret this as an operator statement in the bulk effective field theory, since we used the constraint equation but not the dynamical bulk equations $E^i_j=0$.

\subsection{Excited states and energy levels}\label{subsec:appexcited}
 
In order to analyze excited states, we will consider a metric of the form
\be\label{metric}
ds^2=dw^2 + g_{ij}(x, w)dx^i dx^j = dw^2-g(w)^2 dt^2+r(w, t)^2 d\phi^2, ~~~~ \phi=\phi+2\pi\mu\;,
\ee     
which takes into account the time dependence, but for simplicity we work at zero momentum in the $\phi$ direction. We will discuss in some more detail the algebraic approach followed in the main text, and we will work out a differential equation for the energy on the gravity side, generalizing the analyses in \cite{McGough:2016lol, Kraus:2018xrn}.

The equations of motion are
\bea\label{eq:appEOM}
&&E^t_t = 0 \;\Rightarrow\; \frac{\partial_w^2 r(w,t)}{r(w,t)}= \frac{\eta}{\ell^2}\nonumber \\
&&E^\phi_\phi = 0 \;\Rightarrow\; \frac{\partial_w^2 g(w)}{g(w)}= \frac{\eta}{\ell^2}\nonumber\nonumber \\
&&E^w_w = 0 \;\Rightarrow\; \frac{\partial_w g(w)}{g(w)}\frac{\partial_w r(w,t)}{r(w,t)}-\frac{1}{g(w)^2} \, \frac{\partial_t^2 r(w,t)}{r(w,t)}= \frac{\eta}{\ell^2}\,,
\eea
and the quasilocal stress-energy tensor (\ref{bCTT}) reads
\bea\label{KsTs}
K_{tt}&=&-g g' \nonumber \\
K_{\phi\phi}&=&r\partial_w r \nonumber \\
K &=& g^{ij}K_{ij}=\frac{g'}{g}+\frac{\partial_w r}{r}\\
T_{tt} &=& {-}\frac{g^2}{8\pi G\ell}\left(1-\frac{\ell\partial_w  r}{r} \right)\nonumber \\
T_{\phi\phi} &=& \frac{r^2}{8\pi G\ell}\left(1-\frac{\ell g'}{g} \right) \nonumber\,.
\eea

Let us focus on de Sitter, $\eta=-1$, and choose a constant curvature slicing, denoting the curvature by
\be
2 \frac{\partial_t^2 r(w,t)}{r(w,t)} \equiv \frac{2}{r_0^2}\,, 
\ee
By solving the first two Einstein equations and plugging into the third, the solution is found to be
\be
g(w) = \frac{\ell}{r_0}\,\sin(\frac{w}{\ell})\;\;,\;\;r(w,t) =\frac{L_0}{2\pi}\,g(w) (e^{t/r_0}+\alpha_r e^{-t/r_0})\,,
\ee
where $L_0, \alpha_r$ are arbitrary constants.\footnote{
The case of vanishing curvature has to be solved separately, as the previous set of equations becomes degenerate. The solutions for this case are  AdS and dS sliced by flat space.} In particular, $L_0$ is a parameter with units of length that can be used to change the length of the circle
\be
L(w,t) = 2 \pi r(w,t)\,,
\ee
while keeping the curvature radius $r_0$ as well as $\ell$, fixed. This gives a deficit angle for $\phi$.

The components of the stress tensor become
\be\label{eq:appT00}
T^t_t = T^\phi_\phi = \frac{1}{8\pi G \ell} \left(1-\frac{1}{\tan(\frac{w_c}{\ell})} \right)\,.
\ee
Rewriting $\tan (w_c/\ell)$  in (\ref{eq:appT00}) as a function of the Ricci scalar of the $w=w_c$ surface,
\be
{\mathcal R^{(2)}}= \frac{2}{g(w_c)^2 r_0^2}
\ee
obtains
\be
T^t_t = T^\phi_\phi = \frac{1}{8\pi G \ell} \left(1-\sqrt{\frac{1}{2}\ell^2{\mathcal R^{(2)}}-1}\,\right)\,.
\ee
Therefore, (\ref{Ttautauanswertext}) applies to the class of metrics (\ref{metric}). Note that if we vary the length of the circle while keeping the other parameters fixed (i.e. we only vary $L_0$), we have
\be
E =- L T^t_t\;,\;\partial_L E = -T^t_t =- T^\phi_\phi\,,
\ee
so in our curved slicing we retain the meaning of pressure, $T^\phi_\phi= -\partial_L E$.

Finally, let us show how to extend the approach of \cite{McGough:2016lol, Kraus:2018xrn} to our case. Following \cite{Kraus:2018xrn} (but using our standard convention for $T_{ij}$ as noted above) let us define
\be\label{eps}
{\cal E}=EL = {L}\int d\phi\sqrt{g_{\phi\phi}} u^i u^j T_{ij}
\ee
where $u$ is a timelike unit vector, for us simply $u^t=1/g$. Given (\ref{KsTs}), this becomes
\be\label{er}
{\cal E}={-} \frac{L^2}{8\pi G\ell}\left(1-\frac{\ell\partial_w  r}{r} \right) \,.
\ee
The Einstein equation $E^t_t=0$ is equivalent to
\be\label{diffeq}
\partial_w\left(\frac{\partial_w r}{r} \right)+ \left(\frac{\partial_w r}{r} \right)^2-\frac{\eta}{\ell^2}=0.
\ee
Using (\ref{er}) we can trade $\ell\partial_w r/r$ for ${\cal E}$, and use 
(\ref{eq:identify}), obtaining the equation
\be\label{Ediffeq}
\ell\partial_w \left(\frac{\pi\lambda}{L^2}{\cal E}\right)+\left(1+\frac{\pi \lambda}{L^2}{\cal E}\right)^2=\eta\,.
\ee

This can be processed further, by noting that the variation of $w$ changes $L$ and ${\mathcal R}^{(2)}$, while $\lambda$ remains fixed. Then
\be\label{eq:appint}
\partial_w {\cal E} = \frac{\partial_w L}{L}\, L \partial_L {\cal E} +\frac{\partial_w {\mathcal R}^{(2)}}{{\mathcal R}^{(2)}}\, {\mathcal R}^{(2)} \partial_{{\mathcal R}^{(2)}} {\cal E }\,.
\ee
Here $\partial_w L/L$ can be rewritten in terms of $\cal E$ by use of (\ref{er}); also, $\partial_w {\mathcal R}^{(2)}/{\mathcal R}^{(2)}=-2 \partial_w g/g$, which can be traded by $\partial_w r/r$ and $\mathcal R^{(2)}$ if we use the $E^w_w=0$ equation in (\ref{eq:appEOM}). The result is
\be\label{eq:appint2}
\ell \partial_w {\cal E} =\left(1+\frac{\pi \lambda}{L^2} {\cal E} \right) L \partial_L {\cal E}-2\, \frac{\eta+\frac{1}{2}\ell^2 {\cal R}^{(2)}}{1+\frac{\pi \lambda}{L^2} {\cal E}} {\mathcal R}^{(2)} \partial_{{\mathcal R}^{(2)}} {\cal E }\,.
\ee
Introducing the dimensionless combination $y \equiv \lambda/L^2$ and plugging (\ref{eq:appint2}) into (\ref{eq:appint}), we arrive to
\be
(1+\pi y\, {\cal E}) \,y \partial_y {\cal E}+ \frac{\eta+\frac{1}{2}\ell^2 {\cal R}^{(2)}}{1+\pi y\, {\cal E}} \,{\mathcal R}^{(2)} \partial_{{\mathcal R}^{(2)}} {\cal E }+\frac{\pi}{2}\,y\,{\cal E}^2 =\frac{1-\eta}{2\pi y}\,.
\ee
We conclude that in the case of nonzero curvature, the Einstein equation $E^t_t=0$ gives rise to a partial differential equation for the energy levels, which includes both $L$ and ${\cal R}^{(2)}$ variations.

\bibliography{dS}{}

\providecommand{\href}[2]{#2}\begingroup\raggedright\begin{thebibliography}{10}

\bibitem{Anninos:2012qw}
D.~Anninos, ``{De Sitter Musings},''
  \href{http://dx.doi.org/10.1142/S0217751X1230013X}{{\em Int. J. Mod. Phys.}
  {\bfseries A27} (2012) 1230013},
\href{http://arxiv.org/abs/1205.3855}{{\ttfamily arXiv:1205.3855 [hep-th]}}.

\bibitem{Strominger:2001pn}
A.~Strominger, ``{The dS / CFT correspondence},''
  \href{http://dx.doi.org/10.1088/1126-6708/2001/10/034}{{\em JHEP} {\bfseries
  10} (2001) 034},
\href{http://arxiv.org/abs/hep-th/0106113}{{\ttfamily arXiv:hep-th/0106113
  [hep-th]}}.

\bibitem{Strominger:2001gp}
A.~Strominger, ``{Inflation and the dS / CFT correspondence},''
  \href{http://dx.doi.org/10.1088/1126-6708/2001/11/049}{{\em JHEP} {\bfseries
  11} (2001) 049},
\href{http://arxiv.org/abs/hep-th/0110087}{{\ttfamily arXiv:hep-th/0110087
  [hep-th]}}.

\bibitem{Alishahiha:2004md}
M.~Alishahiha, A.~Karch, E.~Silverstein, and D.~Tong, ``{The dS/dS
  correspondence},'' \href{http://dx.doi.org/10.1063/1.1848341}{{\em AIP Conf.
  Proc.} {\bfseries 743} (2005) 393--409},
  \href{http://arxiv.org/abs/hep-th/0407125}{{\ttfamily arXiv:hep-th/0407125
  [hep-th]}}.
[,393(2004)].

\bibitem{Alishahiha:2005dj}
M.~Alishahiha, A.~Karch, and E.~Silverstein, ``{Hologravity},''
  \href{http://dx.doi.org/10.1088/1126-6708/2005/06/028}{{\em JHEP} {\bfseries
  06} (2005) 028},
\href{http://arxiv.org/abs/hep-th/0504056}{{\ttfamily arXiv:hep-th/0504056
  [hep-th]}}.

\bibitem{Freivogel:2006xu}
B.~Freivogel, Y.~Sekino, L.~Susskind, and C.-P. Yeh, ``{A Holographic framework
  for eternal inflation},''
  \href{http://dx.doi.org/10.1103/PhysRevD.74.086003}{{\em Phys. Rev.}
  {\bfseries D74} (2006) 086003},
\href{http://arxiv.org/abs/hep-th/0606204}{{\ttfamily arXiv:hep-th/0606204
  [hep-th]}}.

\bibitem{Dong:2010pm}
X.~Dong, B.~Horn, E.~Silverstein, and G.~Torroba, ``{Micromanaging de Sitter
  holography},'' \href{http://dx.doi.org/10.1088/0264-9381/27/24/245020}{{\em
  Class. Quant. Grav.} {\bfseries 27} (2010) 245020},
\href{http://arxiv.org/abs/1005.5403}{{\ttfamily arXiv:1005.5403 [hep-th]}}.

\bibitem{Dong:2011uf}
X.~Dong, B.~Horn, S.~Matsuura, E.~Silverstein, and G.~Torroba, ``{FRW solutions
  and holography from uplifted AdS/CFT},''
  \href{http://dx.doi.org/10.1103/PhysRevD.85.104035}{{\em Phys. Rev.}
  {\bfseries D85} (2012) 104035},
\href{http://arxiv.org/abs/1108.5732}{{\ttfamily arXiv:1108.5732 [hep-th]}}.

\bibitem{Anninos:2011ui}
D.~Anninos, T.~Hartman, and A.~Strominger, ``{Higher Spin Realization of the
  dS/CFT Correspondence},''
  \href{http://dx.doi.org/10.1088/1361-6382/34/1/015009}{{\em Class. Quant.
  Grav.} {\bfseries 34} no.~1, (2017) 015009},
\href{http://arxiv.org/abs/1108.5735}{{\ttfamily arXiv:1108.5735 [hep-th]}}.

\bibitem{Dong:2012afa}
X.~Dong, B.~Horn, E.~Silverstein, and G.~Torroba, ``{Moduli Stabilization and
  the Holographic RG for AdS and dS},''
  \href{http://dx.doi.org/10.1007/JHEP06(2013)089}{{\em JHEP} {\bfseries 06}
  (2013) 089},
\href{http://arxiv.org/abs/1209.5392}{{\ttfamily arXiv:1209.5392 [hep-th]}}.

\bibitem{Dong:2018cuv}
X.~Dong, E.~Silverstein, and G.~Torroba, ``{De Sitter Holography and
  Entanglement Entropy},''
  \href{http://dx.doi.org/10.1007/JHEP07(2018)050}{{\em JHEP} {\bfseries 07}
  (2018) 050},
\href{http://arxiv.org/abs/1804.08623}{{\ttfamily arXiv:1804.08623 [hep-th]}}.

\bibitem{Miyaji:2015yva}
M.~Miyaji and T.~Takayanagi, ``{Surface/State Correspondence as a Generalized
  Holography},'' \href{http://dx.doi.org/10.1093/ptep/ptv089}{{\em PTEP}
  {\bfseries 2015} no.~7, (2015) 073B03},
\href{http://arxiv.org/abs/1503.03542}{{\ttfamily arXiv:1503.03542 [hep-th]}}.

\bibitem{Nomura:2018kji}
Y.~Nomura, P.~Rath, and N.~Salzetta, ``{Pulling the Boundary into the Bulk},''
  \href{http://dx.doi.org/10.1103/PhysRevD.98.026010}{{\em Phys. Rev.}
  {\bfseries D98} no.~2, (2018) 026010},
\href{http://arxiv.org/abs/1805.00523}{{\ttfamily arXiv:1805.00523 [hep-th]}}.

\bibitem{Maldacena:2002vr}
J.~M. Maldacena, ``{Non-Gaussian features of primordial fluctuations in single
  field inflationary models},''
  \href{http://dx.doi.org/10.1088/1126-6708/2003/05/013}{{\em JHEP} {\bfseries
  05} (2003) 013},
\href{http://arxiv.org/abs/astro-ph/0210603}{{\ttfamily arXiv:astro-ph/0210603
  [astro-ph]}}.

\bibitem{Harlow:2011ke}
D.~Harlow and D.~Stanford, ``{Operator Dictionaries and Wave Functions in
  AdS/CFT and dS/CFT},''
\href{http://arxiv.org/abs/1104.2621}{{\ttfamily arXiv:1104.2621 [hep-th]}}.

\bibitem{Zamolodchikov:2004ce}
A.~B. Zamolodchikov, ``{Expectation value of composite field T anti-T in
  two-dimensional quantum field theory},''
\href{http://arxiv.org/abs/hep-th/0401146}{{\ttfamily arXiv:hep-th/0401146
  [hep-th]}}.

\bibitem{Smirnov:2016lqw}
F.~A. Smirnov and A.~B. Zamolodchikov, ``{On space of integrable quantum field
  theories},'' \href{http://dx.doi.org/10.1016/j.nuclphysb.2016.12.014}{{\em
  Nucl. Phys.} {\bfseries B915} (2017) 363--383},
\href{http://arxiv.org/abs/1608.05499}{{\ttfamily arXiv:1608.05499 [hep-th]}}.

\bibitem{Cavaglia:2016oda}
A.~Cavaglia, S.~Negro, I.~M. Szecsenyi, and R.~Tateo, ``{$T \bar{T}$-deformed
  2D Quantum Field Theories},''
  \href{http://dx.doi.org/10.1007/JHEP10(2016)112}{{\em JHEP} {\bfseries 10}
  (2016) 112},
\href{http://arxiv.org/abs/1608.05534}{{\ttfamily arXiv:1608.05534 [hep-th]}}.

\bibitem{Dubovsky:2012wk}
S.~Dubovsky, R.~Flauger, and V.~Gorbenko, ``{Solving the Simplest Theory of
  Quantum Gravity},'' \href{http://dx.doi.org/10.1007/JHEP09(2012)133}{{\em
  JHEP} {\bfseries 09} (2012) 133},
\href{http://arxiv.org/abs/1205.6805}{{\ttfamily arXiv:1205.6805 [hep-th]}}.

\bibitem{McGough:2016lol}
L.~McGough, M.~Mezei, and H.~Verlinde, ``{Moving the CFT into the bulk with $
  T\overline{T} $},'' \href{http://dx.doi.org/10.1007/JHEP04(2018)010}{{\em
  JHEP} {\bfseries 04} (2018) 010},
\href{http://arxiv.org/abs/1611.03470}{{\ttfamily arXiv:1611.03470 [hep-th]}}.

\bibitem{Dubovsky:2017cnj}
S.~Dubovsky, V.~Gorbenko, and M.~Mirbabayi, ``{Asymptotic fragility, near
  AdS$_{2}$ holography and $ T\overline{T} $},''
  \href{http://dx.doi.org/10.1007/JHEP09(2017)136}{{\em JHEP} {\bfseries 09}
  (2017) 136},
\href{http://arxiv.org/abs/1706.06604}{{\ttfamily arXiv:1706.06604 [hep-th]}}.

\bibitem{Cardy:2018sdv}
J.~Cardy, ``{The $T\overline T$ deformation of quantum field theory as random
  geometry},''
\href{http://arxiv.org/abs/1801.06895}{{\ttfamily arXiv:1801.06895 [hep-th]}}.

\bibitem{Cottrell:2018skz}
W.~Cottrell and A.~Hashimoto, ``{Comments on $T \bar T$ double trace
  deformations and boundary conditions},''
\href{http://arxiv.org/abs/1801.09708}{{\ttfamily arXiv:1801.09708 [hep-th]}}.

\bibitem{Kraus:2018xrn}
P.~Kraus, J.~Liu, and D.~Marolf, ``{Cutoff AdS$_{3}$ versus the $ T\overline{T}
  $ deformation},'' \href{http://dx.doi.org/10.1007/JHEP07(2018)027}{{\em JHEP}
  {\bfseries 07} (2018) 027},
\href{http://arxiv.org/abs/1801.02714}{{\ttfamily arXiv:1801.02714 [hep-th]}}.

\bibitem{Donnelly:2018bef}
W.~Donnelly and V.~Shyam, ``{Entanglement entropy and $T \overline{T}$
  deformation},''
\href{http://arxiv.org/abs/1806.07444}{{\ttfamily arXiv:1806.07444 [hep-th]}}.

\bibitem{Hartman:2018tkw}
T.~Hartman, J.~Kruthoff, E.~Shaghoulian, and A.~Tajdini, ``{Holography at
  finite cutoff with a $T^2$ deformation},''
\href{http://arxiv.org/abs/1807.11401}{{\ttfamily arXiv:1807.11401 [hep-th]}}.

\bibitem{Taylor:2018xcy}
M.~Taylor, ``{TT deformations in general dimensions},''
\href{http://arxiv.org/abs/1805.10287}{{\ttfamily arXiv:1805.10287 [hep-th]}}.

\bibitem{Bonelli:2018kik}
G.~Bonelli, N.~Doroud, and M.~Zhu, ``{$T \bar{T}$-deformations in closed
  form},'' \href{http://dx.doi.org/10.1007/JHEP06(2018)149}{{\em JHEP}
  {\bfseries 06} (2018) 149},
\href{http://arxiv.org/abs/1804.10967}{{\ttfamily arXiv:1804.10967 [hep-th]}}.

\bibitem{Aharony:2018vux}
O.~Aharony and T.~Vaknin, ``{The TT* deformation at large central charge},''
  \href{http://dx.doi.org/10.1007/JHEP05(2018)166}{{\em JHEP} {\bfseries 05}
  (2018) 166},
\href{http://arxiv.org/abs/1803.00100}{{\ttfamily arXiv:1803.00100 [hep-th]}}.

\bibitem{Aharony:2018bad}
O.~Aharony, S.~Datta, A.~Giveon, Y.~Jiang, and D.~Kutasov, ``{Modular
  invariance and uniqueness of $T\bar{T}$ deformed CFT},''
\href{http://arxiv.org/abs/1808.02492}{{\ttfamily arXiv:1808.02492 [hep-th]}}.

\bibitem{Datta:2018thy}
S.~Datta and Y.~Jiang, ``{$T\bar{T}$ deformed partition functions},''
  \href{http://dx.doi.org/10.1007/JHEP08(2018)106}{{\em JHEP} {\bfseries 08}
  (2018) 106},
\href{http://arxiv.org/abs/1806.07426}{{\ttfamily arXiv:1806.07426 [hep-th]}}.

\bibitem{Giveon:2017nie}
A.~Giveon, N.~Itzhaki, and D.~Kutasov, ``{$ \mathrm{T}\overline{\mathrm{T}} $
  and LST},'' \href{http://dx.doi.org/10.1007/JHEP07(2017)122}{{\em JHEP}
  {\bfseries 07} (2017) 122},
\href{http://arxiv.org/abs/1701.05576}{{\ttfamily arXiv:1701.05576 [hep-th]}}.

\bibitem{Guica:2017lia}
M.~Guica, ``{An integrable Lorentz-breaking deformation of two-dimensional
  CFTs},''
\href{http://arxiv.org/abs/1710.08415}{{\ttfamily arXiv:1710.08415 [hep-th]}}.

\bibitem{Baggio:2018rpv}
M.~Baggio, A.~Sfondrini, G.~Tartaglino-Mazzucchelli, and H.~Walsh, ``{On
  $T\bar{T}$ deformations and supersymmetry},''
\href{http://arxiv.org/abs/1811.00533}{{\ttfamily arXiv:1811.00533 [hep-th]}}.

\bibitem{Chang:2018dge}
C.-K. Chang, C.~Ferko, and S.~Sethi, ``{Supersymmetry and $T \overline{T}$
  Deformations},''
\href{http://arxiv.org/abs/1811.01895}{{\ttfamily arXiv:1811.01895 [hep-th]}}.

\bibitem{usnext}
W.~in~progress~with Jeremias Aguilera~Damia and D.~Freedman.

\bibitem{Balasubramanian:1999re}
V.~Balasubramanian and P.~Kraus, ``{A Stress tensor for Anti-de Sitter
  gravity},'' \href{http://dx.doi.org/10.1007/s002200050764}{{\em Commun. Math.
  Phys.} {\bfseries 208} (1999) 413--428},
\href{http://arxiv.org/abs/hep-th/9902121}{{\ttfamily arXiv:hep-th/9902121
  [hep-th]}}.

\bibitem{Dubovsky:2018bmo}
S.~Dubovsky, V.~Gorbenko, and G.~Hernandez-Chifflet, ``{$ T\overline{T} $
  partition function from topological gravity},''
  \href{http://dx.doi.org/10.1007/JHEP09(2018)158}{{\em JHEP} {\bfseries 09}
  (2018) 158},
\href{http://arxiv.org/abs/1805.07386}{{\ttfamily arXiv:1805.07386 [hep-th]}}.

\bibitem{Deser:1983nh}
S.~Deser and R.~Jackiw, ``{Three-Dimensional Cosmological Gravity: Dynamics of
  Constant Curvature},''
\href{http://dx.doi.org/10.1016/0003-4916(84)90025-3}{{\em Annals Phys.}
  {\bfseries 153} (1984) 405--416}.

\bibitem{Bousso:2001mw}
R.~Bousso, A.~Maloney, and A.~Strominger, ``{Conformal vacua and entropy in de
  Sitter space},'' \href{http://dx.doi.org/10.1103/PhysRevD.65.104039}{{\em
  Phys. Rev.} {\bfseries D65} (2002) 104039},
\href{http://arxiv.org/abs/hep-th/0112218}{{\ttfamily arXiv:hep-th/0112218
  [hep-th]}}.

\bibitem{Martinec:1998wm}
E.~J. Martinec, ``{Conformal field theory, geometry, and entropy},''
\href{http://arxiv.org/abs/hep-th/9809021}{{\ttfamily arXiv:hep-th/9809021
  [hep-th]}}.

\bibitem{Marolf:2012dr}
D.~Marolf and M.~Rangamani, ``{Causality and the AdS Dirichlet problem},''
  \href{http://dx.doi.org/10.1007/JHEP04(2012)035}{{\em JHEP} {\bfseries 04}
  (2012) 035},
\href{http://arxiv.org/abs/1201.1233}{{\ttfamily arXiv:1201.1233 [hep-th]}}.

\bibitem{Maldacena:1997re}
J.~M. Maldacena, ``{The Large N limit of superconformal field theories and
  supergravity},'' \href{http://dx.doi.org/10.1023/A:1026654312961,
  10.4310/ATMP.1998.v2.n2.a1}{{\em Int. J. Theor. Phys.} {\bfseries 38} (1999)
  1113--1133}, \href{http://arxiv.org/abs/hep-th/9711200}{{\ttfamily
  arXiv:hep-th/9711200 [hep-th]}}.
[Adv. Theor. Math. Phys.2,231(1998)].

\bibitem{Susskind:1998dq}
L.~Susskind and E.~Witten, ``{The Holographic bound in anti-de Sitter space},''
\href{http://arxiv.org/abs/hep-th/9805114}{{\ttfamily arXiv:hep-th/9805114
  [hep-th]}}.

\bibitem{Balasubramanian:2014sra}
V.~Balasubramanian, B.~D. Chowdhury, B.~Czech, and J.~de~Boer, ``{Entwinement
  and the emergence of spacetime},''
  \href{http://dx.doi.org/10.1007/JHEP01(2015)048}{{\em JHEP} {\bfseries 01}
  (2015) 048},
\href{http://arxiv.org/abs/1406.5859}{{\ttfamily arXiv:1406.5859 [hep-th]}}.

\bibitem{Balasubramanian:2016xho}
V.~Balasubramanian, A.~Bernamonti, B.~Craps, T.~De~Jonckheere, and F.~Galli,
  ``{Entwinement in discretely gauged theories},''
  \href{http://dx.doi.org/10.1007/JHEP12(2016)094}{{\em JHEP} {\bfseries 12}
  (2016) 094},
\href{http://arxiv.org/abs/1609.03991}{{\ttfamily arXiv:1609.03991 [hep-th]}}.

\bibitem{Balasubramanian:2018ajb}
V.~Balasubramanian, B.~Craps, T.~De~Jonckheere, and G.~Sarosi, ``{Entanglement
  versus entwinement in symmetric product orbifolds},''
\href{http://arxiv.org/abs/1806.02871}{{\ttfamily arXiv:1806.02871 [hep-th]}}.

\bibitem{Rangamani:2009xk}
M.~Rangamani, ``{Gravity and Hydrodynamics: Lectures on the fluid-gravity
  correspondence},''
  \href{http://dx.doi.org/10.1088/0264-9381/26/22/224003}{{\em Class. Quant.
  Grav.} {\bfseries 26} (2009) 224003},
\href{http://arxiv.org/abs/0905.4352}{{\ttfamily arXiv:0905.4352 [hep-th]}}.

\bibitem{Ryu:2006bv}
S.~Ryu and T.~Takayanagi, ``{Holographic derivation of entanglement entropy
  from AdS/CFT},'' \href{http://dx.doi.org/10.1103/PhysRevLett.96.181602}{{\em
  Phys. Rev. Lett.} {\bfseries 96} (2006) 181602},
\href{http://arxiv.org/abs/hep-th/0603001}{{\ttfamily arXiv:hep-th/0603001
  [hep-th]}}.

\bibitem{Hubeny:2007xt}
V.~E. Hubeny, M.~Rangamani, and T.~Takayanagi, ``{A Covariant holographic
  entanglement entropy proposal},''
  \href{http://dx.doi.org/10.1088/1126-6708/2007/07/062}{{\em JHEP} {\bfseries
  07} (2007) 062},
\href{http://arxiv.org/abs/0705.0016}{{\ttfamily arXiv:0705.0016 [hep-th]}}.

\bibitem{Casini:2012ei}
H.~Casini and M.~Huerta, ``{On the RG running of the entanglement entropy of a
  circle},'' \href{http://dx.doi.org/10.1103/PhysRevD.85.125016}{{\em Phys.
  Rev.} {\bfseries D85} (2012) 125016},
\href{http://arxiv.org/abs/1202.5650}{{\ttfamily arXiv:1202.5650 [hep-th]}}.

\bibitem{Aharony:2001pa}
O.~Aharony, M.~Berkooz, and E.~Silverstein, ``{Multiple trace operators and
  nonlocal string theories},''
  \href{http://dx.doi.org/10.1088/1126-6708/2001/08/006}{{\em JHEP} {\bfseries
  08} (2001) 006},
\href{http://arxiv.org/abs/hep-th/0105309}{{\ttfamily arXiv:hep-th/0105309
  [hep-th]}}.

\bibitem{Brattan:2011my}
D.~Brattan, J.~Camps, R.~Loganayagam, and M.~Rangamani, ``{CFT dual of the AdS
  Dirichlet problem : Fluid/Gravity on cut-off surfaces},''
  \href{http://dx.doi.org/10.1007/JHEP12(2011)090}{{\em JHEP} {\bfseries 12}
  (2011) 090},
\href{http://arxiv.org/abs/1106.2577}{{\ttfamily arXiv:1106.2577 [hep-th]}}.

\bibitem{Anninos:2017hhn}
D.~Anninos and D.~M. Hofman, ``{Infrared Realization of dS$_2$ in AdS$_2$},''
  \href{http://dx.doi.org/10.1088/1361-6382/aab143}{{\em Class. Quant. Grav.}
  {\bfseries 35} no.~8, (2018) 085003},
\href{http://arxiv.org/abs/1703.04622}{{\ttfamily arXiv:1703.04622 [hep-th]}}.

\end{thebibliography}\endgroup
\bibliographystyle{utphys}

\end{document}